\documentclass[aip,amssymb,reprint,groupedaddress,floatfix]{revtex4-2}
\usepackage{graphicx}
\usepackage{amsmath}
\usepackage{hyperref}
\usepackage{physics}
\usepackage{dcolumn}
\usepackage{bm}
\usepackage{braket}
\usepackage[dvipsnames]{xcolor}
\usepackage{changes}

\begin{document}

\title{Quantum thermodynamics of driven-dissipative condensates}

\author{Lu\' \i sa Toledo Tude and Paul R. Eastham}

\affiliation{School of Physics, Trinity College Dublin, Dublin 2, Ireland}
\affiliation{Trinity Quantum Alliance, Unit 16, Trinity Technology and Enterprise Centre, Pearse Street, Dublin 2, Ireland}
\date{\today}

\begin{abstract}
Polariton condensates occur away from thermal equilibrium, in an open system where heat and particles are continually exchanged with  reservoirs. These phenomena have been extensively analyzed in terms of kinetic equations. Based on the collection of knowledge about polariton kinetics provided by these simulations and by experimental works, we constructed a  few-level model that captures the main processes involved in the buildup of a ground-state population of polaritons.
This allows condensation to be understood as the output of a thermal machine and exposes the thermodynamic constraints on its occurrence. The model consists of a three-level system interacting with a field and connected to a hot and a cold thermal reservoir that represent a non-resonant pump and the lattice phonons. This subsystem can drive a condensate, through polariton-polariton scattering, which produces work in the form of coherent light emission from the microcavity. We obtain a phase diagram as a function of the temperatures of the two baths and investigate the possible types of phase transition that lead to the condensate phase.
\end{abstract}


\maketitle



\section{Introduction}

Bose-Einstein condensates are ordered states which show macroscopic quantum effects like superfluidity. While originally an equilibrium phenomenon, there are now many situations where condensation occurs in the non-equilibrium steady-states of driven open systems. Examples include condensates of exciton-polaritons in inorganic~\cite{keeling_boseeinstein_2020} and organic semiconductors~\cite{deng_exciton-polariton_2010}, photons in dye-filled cavities~\cite{klaers_boseeinstein_2010,kirton_nonequilibrium_2013}, plasmon-polaritons~\cite{hakala_boseeinstein_2018}, and magnons~\cite{demokritov_boseeinstein_2006}. These particles have a finite lifetime, and the condensate must be maintained against losses by gain from an external pump. 


Some condensates, such as those of photons and magnons, quickly reach a quasi-equilibrium state, and so can to an extent be treated using equilibrium thermodynamics. For others, however, this description does not apply. Progress has been made by studying the dynamics and steady-states of different models, using kinetic equations and field-theoretic approaches, among other techniques. This diversity of approaches makes it difficult to identify the general requirements for condensation. Here, we suggest a universal description of non-equilibrium condensation as a heat engine. We show how condensation can be described using a three-level laser model, whose connection to the thermodynamics of heat engines, first pointed out by Scovil and Schulz-DuBois, has been extensively studied~\cite{scovil_three-level_1959,geusic_quantum_1967,kosloff_quantum_1984,geva_three-level_1994,geva_quantum_1996,mitchison_quantum_2019}. For definiteness, we focus on the exciton-polariton condensate, and use the collection of knowledge provided by theoretical and experimental works to construct the three-level model. 

Our approach restricts condensation into a minimal model allowing the extraction of fundamental thermodynamic constraints of the system, such as the Carnot limit. We investigate properties such as efficiency, and determine the conditions required for condensation. An important result is that the occurrence of condensation is determined by two temperatures whose difference controls the direction of energy flow in the system, governing the formation of the condensate, and allowing the losses to be overcome. Our work further clarifies that non-equilibrium condensation requires a population inversion, albeit of an unconventional kind, and emphasizes the need for an effective coupling to a cold reservoir. This last requirement -- the need for a rapid depopulation of the lower state -- is well known in the context of lasers, but is rarely explicit in the literature on driven-dissipative condensates.

Many previous works on polariton condensation have studied the kinetic equations describing the relaxation of polaritons within their dispersion relation, as well as their generation and decay~\cite{deng_exciton-polariton_2010,kavokin_cavity_2003,tassone_bottleneck_1997,tassone_exciton-exciton_1999,doan_condensation_2005, hartwell_well_2008}. The relaxation was shown to occur through phonon-polariton and polariton-polariton scattering, leading to a non-equilibrium phase transition to a condensate above a critical pump strength. A simplified thermodynamic description has been given by Porras et. al.~\cite{porras_polariton_2002}, in which the polaritons in the high-momentum and energy states were treated as a heat bath. Going beyond the standard rate equation approaches a mean-field analysis of the dynamics has shown that there is an exceptional point conditioning a second phase transition in the system~\cite{hanai2019non}. Another recent work considers the polariton kinetics assuming rapid thermalization, allowing the calculation of the density matrix of the non-equilibrium condensate~\cite{shishkov_exact_2021}. 

The thermodynamics of three-level amplifiers has been studied before, and our main contribution is to apply these results to driven-dissipative condensation. However,  this application does require two extensions of the general framework. Firstly, we consider not just amplification, but the balance of gain and loss in the full system comprising a condensate driven by a gain medium. This allows us to compute non-equilibrium phase diagrams and study how thermodynamic quantities such as efficiency vary across them. Secondly, we allow for the possibility that the thermal machine operates between reservoirs at different chemical potentials, as well as different temperatures, implying that it absorbs work as well as heat. This is necessary in practice for driven-dissipative condensates, as discussed below, but also relevant in cases such as electrically-driven lasers, where the voltage bias is a source of work. 



\section{Condensates as heat engines}

Driven-dissipative condensates are typically understood as open quantum systems, in which energy from an incoherent pump reservoir is converted into a coherent condensate, which is in turn emitted into an environment. As discussed further below, we argue that if the pump is an incoherent source it corresponds to a heat bath, while the coherent emission from the condensate is a source of work. With these identifications we identify the basic form of condensate as a heat engine, converting heat, from the pump, into work. It follows immediately, as a consequence of the second law of thermodynamics, that a consistent description of condensation requires consideration of a cold reservoir, in addition to the hot reservoir representing the pump. Note that in the following we extend these considerations to allow the pump to provide work, as well as heat. We will find that this allows condensation to occur when the thermal machine is not operating as a heat engine, but as a dissipator or refrigerator.

\begin{figure}[t]
    \centering
     \includegraphics[width=1\linewidth]{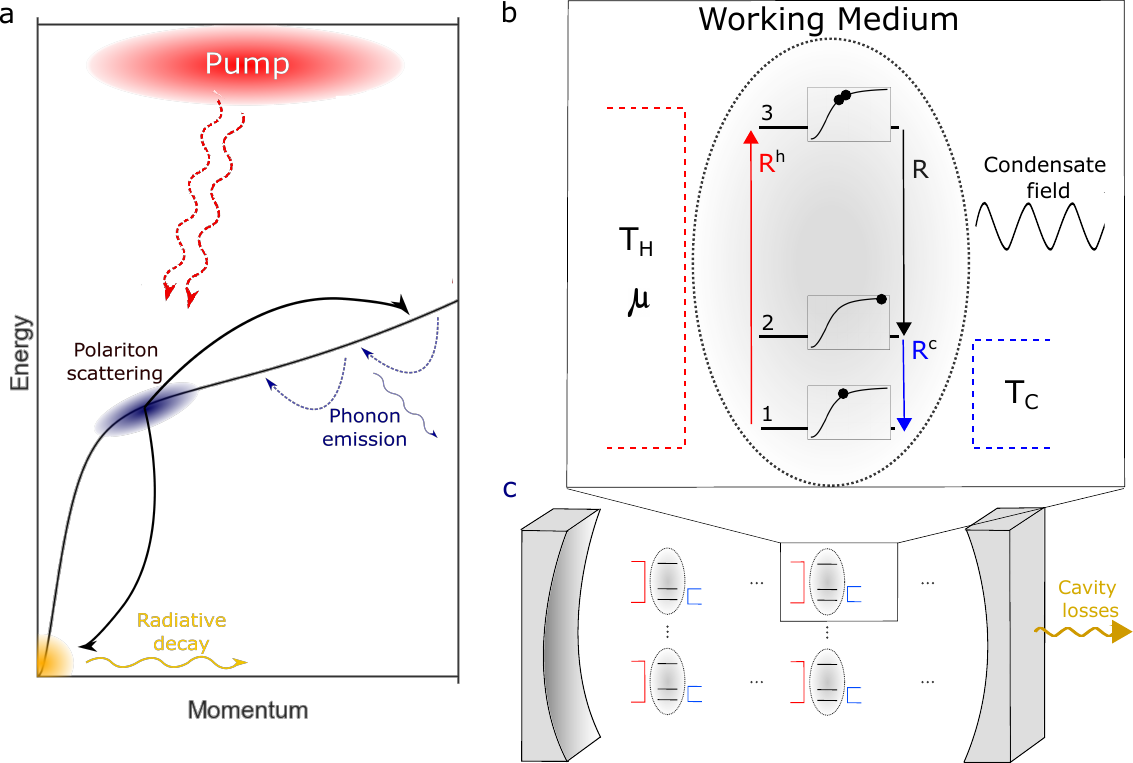}
    \caption{(a) Illustration of the main processes that lead to condensation of inorganic microcavity polaritons. A non-resonant pump creates excitons that relax to lower energy states through phonon emission. Pair-wise polariton scattering leads to occupation in the lower energy modes if it can overcome radiative losses. (b) The condensation process is modeled as the outcome of a three-level heat engine with a hot bath connecting the ground state and the most energetic state and a cold bath connecting the ground state to the middle state. The condensate is a classical field interacting with the two excited states of the working medium. The rectangles on the top of each level show the correspondent states in the dispersion curve. The first and third levels corresponds to $n-1$ and $n$ polaritons in the bottleneck, respectively. The middle level corresponds to a state with a high energy exciton (c) Model of a microcavity comprising $M$ three-level systems that contribute to growth of the condensate, which competes with loss due to the finite polariton lifetime.}
     \label{fig:diagram}
\end{figure}

To develop this idea further we construct a few-level model of condensation. For definiteness, we consider inorganic microcavities, where the processes involved in forming and maintaining the condensate are generally accepted. As illustrated in  Fig.~\ref{fig:diagram}(a), particles are created by a high-energy pump, which produces electron-hole pairs that populate the exciton states at high momentum. This population quickly begins to thermalize with the emission of acoustic phonons~\cite{imamoglu_nonequilibrium_1996}, but such scattering becomes ineffective for the polaritonic final states at low energy and momentum. As confirmed by kinetic simulations~\cite{kasprzak_boseeinstein_2006,balili_bose-einstein_2007,kavokin_cavity_2003,kavokin_microcavities_2017,carusotto_quantum_2013,cao_condensation_2004,doan_condensation_2005,tassone_exciton-exciton_1999,kasprzak_formation_2008,piermarocchi_nonequilibrium_1996,hartwell_numerical_2010,malpuech_polariton_2002,porras_polariton_2002}, this bottleneck effect~\cite{tassone_bottleneck_1997} can be overcome by polariton-polariton scattering.  If the density is large enough this scattering into low-energy states can exceed their loss. It then becomes a stimulated process which leads to condensation~\cite{baumberg_polariton_2002}.

To determine the thermodynamic constraints on condensation we consider a minimal model, consisting of a heat engine whose working medium is a three-level system with energies $e_1$, $e_2$, and $e_3$. It can be argued that this is the simplest possible case~\cite{linden_how_2010}. As shown in Fig.~\ref{fig:diagram}(b), we suppose that the highest energy state of the working medium, $\ket{3}$, corresponds to a population of $n$ polaritons in the bottleneck region. These can undergo pairwise scattering, adding one particle to the condensate and promoting one polariton to a higher-energy exciton state. Due to its macroscopic occupation, such stimulated scattering can be treated as an interaction with a condensate field external to the working medium. This takes the latter to an intermediate state, $\ket{2}$, that has a particle in a high energy state. The high-energy exciton can then transition back to the bottleneck region by phonon emission, leading to the state $\ket{1}$ with bottleneck population $n-1$. In the terminology of the polariton parametric oscillator~\cite{baumberg_polariton_2002}, the populated bottleneck states are the pump states, and the high-energy exciton state generated by pairwise scattering is the idler.


The three-level working medium drives the polariton condensate, which we treat as a classical mean field. The relevant Hamiltonian is \begin{equation}
    H_{s} = g_{c} (p_c^\dag p_i^\dag p_p p_p +  p_c p_i p_p^\dag p_p^\dag ),\label{eq:Hlasernotation}
\end{equation} where $p_c$, $p_p$, and $p_i$ are the annihilation operators for particles in the condensate state, pump state, and idler state, respectively, and $g_c$ is the interaction strength. We consider a macroscopically occupied condensate mode and approximate $g_c p_c\approx g_c \langle p_c \rangle = e^{-i\omega t}\Omega/2$, where $\omega$ is the condensate frequency, and for a condensate of $N$ particles \begin{equation}\Omega= 2 g_c \sqrt{N}.\label{eq:rabicondpop}\end{equation} In the basis of the three-level system, the final three terms in each product in Eq. (\ref{eq:Hlasernotation}) become transition operators $p_i^\dag p_p p_p=\ket{2}\bra{3}$, and the Hamiltonian for the scattering process reduces to that of the three-level system driven by a field corresponding to the condensate,
\begin{equation}
   H_{s} =  \frac{\Omega}{2} \big( e^{-i\omega t} \ket{3}\bra{2} + e^{i\omega t} \ket{2}\bra{3}\big).\label{eq:driventls}
\end{equation} This implies that the energy transfers between the three-level system and the condensate are work, in accordance with the standard partition of energy currents~\cite{adesso_thermodynamics_2018} \begin{equation}
    \frac{d\langle E \rangle}{dt} = \Tr[\rho \frac{d H}{dt}] + \Tr[ \frac{d \rho}{dt} H] = \langle \dot{W} \rangle + \langle \dot{Q} \rangle,\label{eq:1stlaw}
\end{equation} in which work arises from the time-dependence of a Hamiltonian and heat from that of the density-matrix. Using this along with Eq. (\ref{eq:driventls}), the power supplied to the condensate is
\begin{align}
    \langle \dot{W} \rangle = \left\langle\frac{\partial H_{s}}{\partial t} \right\rangle & = \omega \Omega \mathfrak{Im}\left\{\Tr[\rho e^{i\omega t}\ket{2}\bra{3}]\right\} \nonumber \\ & =\omega\Omega\mathfrak{Im} \rho_{32},\label{eq:power}
\end{align} where $\rho_{32}$ is an element of the density matrix in the time-dependent basis introduced below. 


Additionally, the three-level system exchanges energy and particles with a reservoir of acoustic phonons and a reservoir of excitons that is generated by the pump. The phonons act as the cold bath of the heat engine, extracting heat from the working medium as the excitons relax to the bottleneck ($\ket{2} \rightarrow \ket{1}$). The excitons generated by the external pump act as the hot bath. To stay within the framework of a three-level model we suppose they feed the states at the bottleneck, causing transitions from $\ket{1}$ to $\ket{3}$. The lifetime of a reservoir exciton is relatively long, so that the hot reservoir forms a quasi-equilibrium state with temperature $T_h$ and chemical potential $\mu$. This means that the heat engine operates with a chemical potential drop between the hot and cold reservoirs. Since particles flow through the engine across this chemical potential drop, its energy input consists of both work and heat.

\begin{figure}[t]
    \centering
   \includegraphics[width=0.47\textwidth]{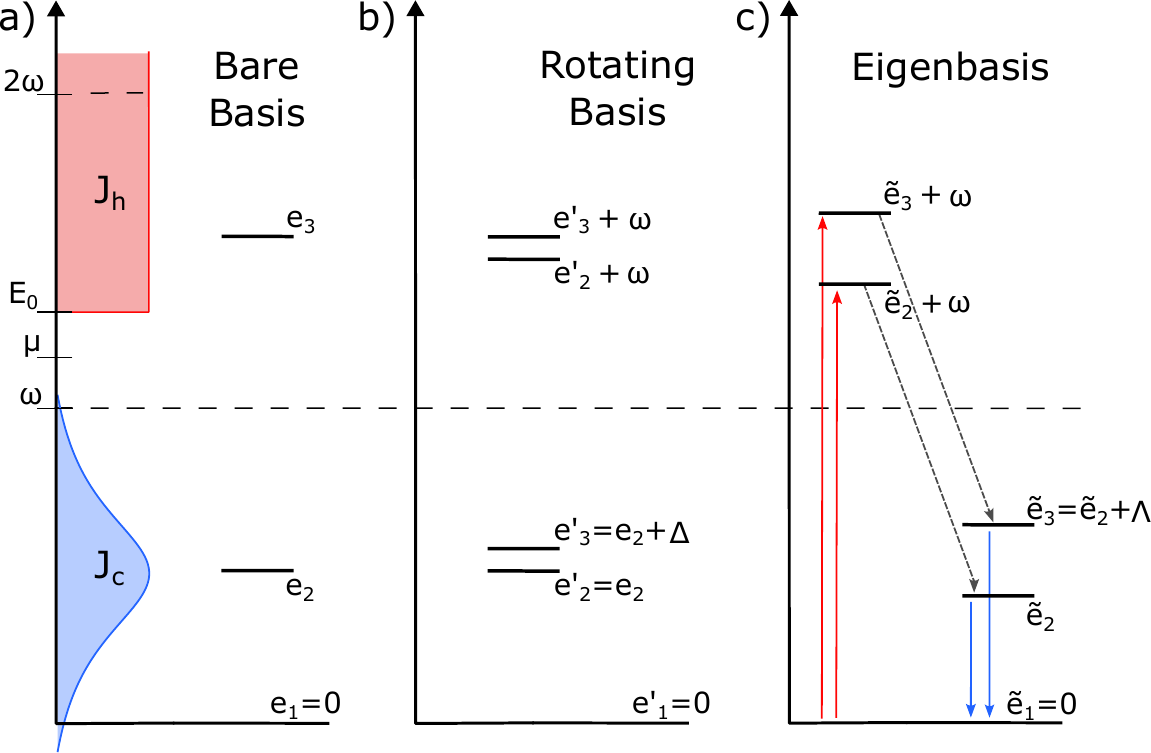}
    \caption{Energy diagram of the three-level system in the three different bases, and the spectral densities of the hot and cold bath. (a) The energy levels, $e_i$, in the bare basis correspond to the ones shown in Fig.~1. (b) When the Hamiltonian is transformed into a rotating basis its time dependence is lost, however, the states acquire a periodicity in energy. Two Floquet zones are included  ($e_i^\prime$) in the diagram. (c) The interaction with the condensate field mixes the states of the rotating basis to produce eigenstates with energies ($\Tilde{e}_2, \Tilde{e}_3$), split by $\Lambda = \sqrt{\Delta^2 +\Omega^2}$. These states are replicated in the two Floquet zones. The red and blue arrows indicate transitions due to the hot and cold baths respectively (see text).}
     \label{fig:dressedstates}
\end{figure}

In summary, the thermal machine operates in a mode in which the hot reservoir inputs work and heat into the three-level system, which performs work by interacting with the condensate and emits the rest of the energy into a cold bath of phonons. Combining the free Hamiltonian of the three-level system, the coupling to the condensate, and the coupling to the baths, we have 
\begin{equation}
    \begin{aligned}
        H_0+H_{sb} =& e_3 \ket{3}\bra{3}+ e_2 \ket{2}\bra{2} \\
        +&   e_1 \ket{1}\bra{1}+ \frac{\Omega}{2} \big(e^{-i \omega t} \ket{3}\bra{2} + e^{i \omega t} \ket{2}\bra{3}\big) \\
        +& \sum_k g_k^{\text{ph}} \big( \ket{1}\bra{2}  + \ket{2}\bra{1}\big) (b_k + b_k^\dag) \\
        +& \sum_q g_q^{\text{p}}   \big( \ket{3}\bra{1} + \ket{1}\bra{3} \big) (x_q + x_q^\dag).
    \end{aligned}\label{eq:Hbare}
\end{equation}
The energies $e_{1,2,3}$, which we refer to as the bare states, are depicted in Fig.~\ref{fig:dressedstates}a. The third and fourth lines of Eq. (\ref{eq:Hbare}) represent the interactions with the cold and hot baths, with coupling strengths $g_k^{\text{ph}}$, and $g_q^{\text{p}}$, and annihilation operators $b_k$ and $x_q$, respectively. In the present case, they refer to phonons, and excitons in states which repopulate the bottleneck. One could also consider, in addition to the process where a pump-generated exciton transfers into a bottleneck state, one in which it transfers into the higher-energy state, which then relaxes to the bottleneck by phonon emission. This could be incorporated by using a four-level model, with an additional state $\ket{4}$, representing the state with an additional high-energy exciton, which then relaxes to $\ket{3}$ by phonon emission.  Such a model could be further extended to include higher phonon states, giving a form similar to that used for photon condensation by Kirton and Keeling~\cite{kirton_nonequilibrium_2013}. However, inclusion of these additional pathways would not be expected to qualitatively affect our results, because under reasonable conditions the reservoir population will decrease with increasing energy, so that the strongest effect of repopulating the bottleneck comes from the process we consider. 

We eliminate the time dependence of Eq. (\ref{eq:Hbare}) by using the rotating frame $(\ket{1_R}, \ket{2_R}, \ket{3_R}) = (\ket{1}, \ket{2}, e^{-i \omega t}\ket{3})$, which leads leads to 
\begin{equation}
    \begin{aligned}
        H_0^\prime+H_{sb}^\prime =& (e_2 + \Delta) \ket{3_R}\bra{3_R}+ e_2 \ket{2_R}\bra{2_R} \\
        +&   e_1 \ket{1_R}\bra{1_R}+ \frac{\Omega}{2} \big( \ket{3_R}\bra{2_R} + \ket{2_R}\bra{3_R}\big) \\
        +& \sum_k g_k^{\text{ph}} \big( \ket{1_R}\bra{2_R}  + \ket{2_R}\bra{1_R}\big) (b_k + b_k^\dag) \\
        +& \sum_q g_q^{\text{p}}   \big( \ket{3_R}\bra{1_R}x_q + \ket{1_R}\bra{3_R}x_q^\dag\big).
    \end{aligned}\label{eq:HR}
\end{equation}
The transformation to this rotating basis would produce a time dependence in the coupling to the hot bath, which, however, has been removed by transforming to an interaction picture with respect to a Hamiltonian $\sum_q \omega x_q^\dag x_q$, and making the rotating wave approximation in the system-reservoir coupling. The detuning $\Delta = (e_{3}- e_2) - \omega$ corresponds to the energy difference between the low-energy final state of the scattering and the condensate. The energy levels given by the first three terms of this Hamiltonian are depicted in Fig.\ref{fig:dressedstates}b. We note that, as the energies of a periodically-driven system, they are defined up to multiples of the driving frequency $\omega$. We define the zero of energy to be $e_1 = 0$. For the other energies we use values representative of a GaAs microcavity, $e_2 = 5$ meV and $e_3 - e_2 = 1$ eV.

\section{Methods}

To analyze Eq. (\ref{eq:HR}) we use the standard method, previously applied to a three-level heat engine in~\cite{geva_three-level_1994}, in which one transforms to the eigenbasis of $H_0^\prime$ and eliminates the heat baths within the Born-Markov approximation. The transformation to the eigenbasis of $H_0^\prime$ is effected by the rotation \begin{equation}\begin{pmatrix} \ket{\Tilde{3}_R} \\ \ket{\Tilde{2}_R} \\ \ket{\Tilde{1}_R}\end{pmatrix}=\begin{pmatrix}\cos \theta/2 & \sin \theta/2 & 0 \\ -\sin \theta/2 & \cos \theta/2 & 0 \\ 0 & 0 & 1\end{pmatrix}\begin{pmatrix} \ket{3_R} \\ \ket{2_R} \\ \ket{1_R}\end{pmatrix}=U \begin{pmatrix} \ket{3_R} \\ \ket{2_R} \\ \ket{1_R}\end{pmatrix},\end{equation} which makes $\Tilde{H}_0 = U^\dagger H_0^\prime U$ diagonal when $\tan\theta =\Omega/\Delta$. The resulting energies, $\Tilde{e}_1= e_1 =0$ and $\Tilde{e}_{2,3}=\big(e_3-\omega + e_2 \mp \sqrt{\Delta^2+\Omega^2}\big)/2$ are depicted in Fig. \ref{fig:dressedstates}c.To derive a master equation for the working medium we follow common practice and neglect the principal value terms, but do not make the secular approximation. The principal value terms emerge from tracing out the bath degrees of freedom and correspond to energy shifts, which can be included in the original Hamiltonian. Further justification for these approximations is given in Ref.~\onlinecite{murphy_laser_2022}. The resulting equations-of-motion are given in Eqs.  (\ref{eq:dmeom11}--\ref{eq:dmeom33}) and Eqs. (\ref{eq:odeom23c}--\ref{eq:odeom12h}).

The method of counting field statistics was implemented to obtain the energy currents to the bath. The procedure consists in re-deriving the master equation introducing a variable, called the counting field, that keeps track of the energy exchanged with the baths~\cite{esposito_nonequilibrium_2009,gasparinetti_heat-exchange_2014}. We find
    \begin{align}
\langle \dot{Q}_\text{c} \rangle&=\Tilde{e}_3 R^c_{3}+\Tilde{e}_2 R^c_{2},\label{eq:Qc}  \\ 
\langle \dot{E}_\text{h} \rangle&=(\Tilde{e}_3+\omega) R^h_3 + (\Tilde{e}_2+\omega) R^h_2. \label{eq:Eh}
\end{align}
The energy current to the cold bath is entirely heat and hence denoted $Q_c$, whereas that to the hot bath comprises both heat and work, with the latter arising from the flow of particles from the exciton reservoir, which in general has a non-zero chemical potential. The rates which appear in the energy currents are 
\begin{align}
    R^{c}_{2,3}&=\pi J_c(\Tilde{e}_{2,3}) [\mp (n_c(\Tilde{e}_{2,3}) +1)  \sin\theta \mathfrak{Re}[\rho_{23}] \nonumber \\ &+ (1\pm \cos\theta)((n_c(\Tilde{e}_{2,3}) +1) \rho_{22} - n_c(\Tilde{e}_{2,3}) \rho_{11})] \label{eq:coldcurbare}
\end{align}   
\begin{align}
    R^{h}_{2,3}=\pi J_h(\Tilde{e}_{2,3}+ &\omega)  [ \mp (n_h(\Tilde{e}_{2,3}+ \omega) +1)  \sin\theta \mathfrak{Re}[\rho_{23}] \nonumber \\ &+ (1\mp\cos\theta)((n_h(\Tilde{e}_{2,3}+ \omega) +1) \rho_{33} \nonumber \\ & \qquad\qquad\qquad - n_h(\Tilde{e}_{2,3}+ \omega) \rho_{11})].\label{eq:hotcurbare}
\end{align}   Here $\rho_{ij}$ are the elements of the density matrix in the basis $(\ket{1_R},\ket{2_R},\ket{3_R})$. These rates can also be expressed in terms of the elements of the density matrix in the diagonal basis, $(\ket{\Tilde{1}_R},\ket{\Tilde{2}_R},\ket{\Tilde{3}_R})$. Denoting those elements by $\Tilde{\rho}_{ij}$ the rates are 

\begin{widetext}

\begin{align}
R^{c}_{2,3}=\pi J_c(\Tilde{e}_{2,3}) \{ & (1\pm \cos\theta)[(n_c(\Tilde{e}_{2,3}) +1) \Tilde{\rho}_{22,33} - n_c(\Tilde{e}_{2,3}) \Tilde{\rho}_{11}]  + (n_c(\Tilde{e}_{2,3}) +1) \sin\theta \mathfrak{Re}[\Tilde{\rho}_{23}]\}
\label{eq:coldcur} \\ 
R^{h}_{2,3}=\pi J_h(\Tilde{e}_{2,3}+ \omega) \{ & (1\mp\cos\theta)[(n_h(\Tilde{e}_{2,3}+ \omega) +1) \Tilde{\rho}_{22,{33}}  - n_h(\Tilde{e}_{2,3}+ \omega) \Tilde{\rho}_{11}] - (n_h(\Tilde{e}_{2,3}+ \omega) +1) \sin\theta \mathfrak{Re}[\Tilde{\rho}_{23}]\}.\label{eq:hotcur}
\end{align}

\end{widetext}

$J_{\text{c,h}}$ are the spectral densities for the cold and hot baths, respectively. In the following we take a Lorentzian for the cold phonon bath, $J_{\text{c}}(x) = (\alpha^\text{c}/2)\left[(x- e_2)^2 + g_{\text{ph}}^2\right]^{-1}$, and a step function for the hot pump bath $J_{\text{h}}(x) = \alpha^\text{h} \Theta (x-E_0)$. The parameters $\alpha^{\text{c,h}}$ control the maximum values of the spectral density, and  $g_{\text{ph}}$ the width of the Lorentzian distribution, whose value needs to be large enough to allow the Born-Markov approximation. $E_0$ is the energy of the lowest exciton state in the pump reservoir. The bath occupation function for the cold bath is $n_\text{c}(E) =(e^{\beta_\text{c} E}-1)^{-1}$, and that for the hot bath is the corresponding grand-canonical form, including the chemical potential, $n_\text{h}(E) = (e^{\beta_\text{h} (E- \mu)}-1)^{-1}$.

\section{Results}


We begin by analyzing the dynamics of the cycle in terms of the occupations of the states $\ket{n_R}$ ($n=1,2,3$). The diagonal elements of the density matrix, in the rotating basis, obey 
\begin{align}
& (\dot{\rho})_{11}=\left(R_2^c+R_3^c\right)+\left(R_2^h+R_3^h\right) \equiv R^c-R^h \label{eq:dmeom11}\\
& (\dot{\rho})_{22}=-\left(R_2^c+R_3^c\right)+R = -R^c+R \label{eq:dmeom22} \\
& (\dot{\rho})_{33}=-\left(R_2^h+R_3^h\right)-R=R^h-R. \label{eq:dmeom33}
\end{align} Here \begin{equation} R=\langle\dot{W}\rangle/\omega=\Omega \mathfrak{Im} \rho_{32},\label{eq:gain}\end{equation} and $R_{2,3}^{c,h}$ are given by Eqs. (\ref{eq:coldcurbare}) and (\ref{eq:hotcurbare}) or (\ref{eq:coldcur}) and (\ref{eq:hotcur}). From these expressions we identify $R^{c}=(R_{2}^c+R_{3}^c)$ as the rate of population transfer from $\ket{2_R}$ to $\ket{1_R}$, due to the cold bath, $R^{h}=-(R_{2}^h+R_{3}^h)$ at that from $\ket{1_R}$ to $\ket{3_R}$, due to the hot bath, and $R$ as that from $\ket{3_R}$ to $\ket{2_R}$, due to the interaction with the condensate. The steady-state condition, \begin{equation} R=R^c=R^h, \label{eq:cyclecond}\end{equation} is that the rates around each part of the cycle are equal.

The energy fluxes to the baths, Eqs. (\ref{eq:Qc},\ref{eq:Eh}), can be interpreted~\cite{geva_quantum_1996} in terms of the eigenstates of $H_0^\prime$, $\ket{\Tilde{n}_R}$, which differ from the states $\ket{n_R}$ around which the population circulates. Due to the oscillating driving field the states are Floquet states associated with a periodic quasi-energy. The transformation to the rotating basis introduces replicas of the original energy levels, which are then mixed by the driving field to form dressed-states with shifted energies. This process is illustrated in Fig.\ \ref{fig:dressedstates}, which shows the original energy levels in the left panel (a), their replicas under periodic driving, (b), and the dressed states obtained by diagonalizing $H_0^\prime$, (c). 

The heat current to the cold bath, Eq. (\ref{eq:Qc}) is the sum of two contributions, each the product of the quasi-energy of a dressed state $\Tilde{e}_2,\Tilde{e}_3$, and one of the two rates, $R^c_{2,3}$. Thus, we can interpret this heat flow as arising from two transitions in the Floquet spectrum, as illustrated in Fig.\ \ref{fig:dressedstates}c. The hot bath acts similarly, but induces transitions from the ground state to levels in the second quasi-energy band. The cycle is closed by the power output, Eq. (\ref{eq:power}), that consists of quanta of energy $\omega$ emitted at the rate $R$. This is illustrated in Fig.\ \ref{fig:dressedstates}, as two independent transitions, each between equivalent dressed states, suggesting the formation of two independent cycles~\cite{geva_quantum_1996}. However, it should be noted that this interpretation implies a secular approximation and only holds in a strong-driving limit. The energy currents cannot in general be interpreted as coming from two separate cycles, as $R^c_{2,3} \neq -R^h_{2,3}$. In fact, the rates in Eqs. (\ref{eq:coldcur},\ref{eq:hotcur}) contain contributions of the form of Boltzmann transition rates, and interference terms weighted by the mixing angle $\theta$. The latter are responsible for coupling the two fictitious cycles, that is,  because of the sign difference they constitute transitions to the ground state in inverted directions, which is effectively an inter-cycle transition. 



The first law of thermodynamics is $\langle \dot{W} \rangle+\langle\dot{Q}_c\rangle+\langle\dot{E}_h\rangle=0$. Using the forms for $\langle \dot{W} \rangle, \langle\dot{Q}_c\rangle$, and $\langle \dot{E}_h\rangle$, and the steady-state condition, Eq. (\ref{eq:cyclecond}), it becomes \begin{equation}
    \Tilde{e}_2 (R^c_2 + R^h_2) + \Tilde{e}_3 (R^c_3 + R^h_3) = 0 .
\end{equation} This condition is not exactly satisfied, but holds to a good approximation in the parameter regimes used here, where the Born-Markov approximation is appropriate.

We can describe a complete system, including the condensate, by noting that the growth of the condensate comes from the output of a large number $M$ of identical three-level heat engines. $M$ represents the number of states in the bottleneck region that generate scattering to the ground state of the polariton distribution, and is on the order of $M=10^4$ in typical cases. The growth rate of a condensate, of population $N$, is then $M R(N)$, where the dependence of the cycle rate $R$ on $\Omega$ and therefore $N$ has been made explicit. This growth competes against the losses due to radiative polariton decay, with rate $\gamma$. The steady-state condition for the condensate number, $N_c$, is then that \begin{equation} M R(N_c)=N_c\gamma.\label{eq:sscond}\end{equation}

 
In the following, we choose parameter values typical of GaAs microcavities. The prefactor of the phonon spectral density is chosen in line with the exciton relaxation times obtained numerically and experimentally~\cite{kavokin_cavity_2003} $\alpha^{\text{c}} = 0.1$ ps$^{-1}$. We take the linewidth of the phonons to be $g_{\text{ph}} =1.7$ ps$^{-1}$. The temperature of the cold bath is the lattice temperature of the semiconductor, which we take to be $10$ K. The spectral density, as well as the chemical potential, and temperature of the hot bath, depend non-trivially on the incoherent pump. We choose $\alpha^{\text{h}} = 0.2$ ps$^{-1}$, $E_0=1$ eV, $E_0 - \mu= 8$ meV and $T_\text{h} = 200$ K initially, and investigate the effects of varying these parameters. We assume a polariton lifetime of $1/\gamma=1$ ps. For the polariton-polariton scattering strength we use the expression derived in~\cite{tassone_exciton-exciton_1999} to estimate $g_c \approx 0.048$ ps$^{-1}$. The detuning appears in our model with fixed condensate and transition energies, and will determine the condensate in-scattering rate, i.e., the gain from the working medium. In practice, however, the condensate, pump, and idler states lie in a continuum, and a range of detunings are present. We expect condensation to occur in the most favorable mode, and so focus on the resonant case $\Delta=0$. 

\begin{figure}[t]
    \centering
     \includegraphics[width=.9\linewidth]{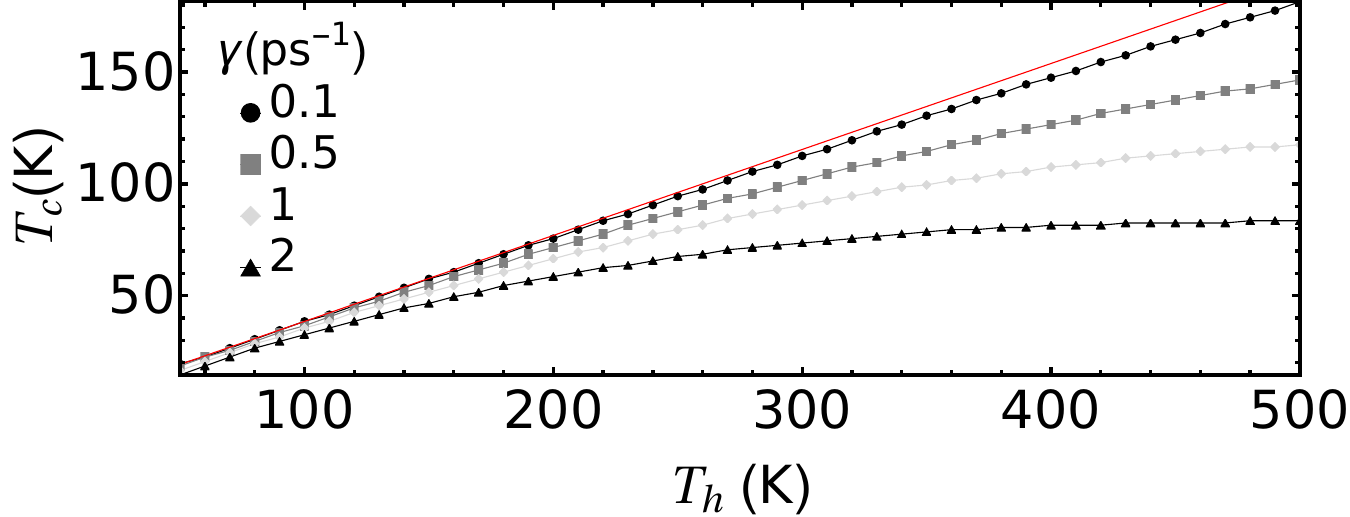}
    \caption{Phase diagram of polariton condensation as a function of the temperatures of the hot and cold reservoir for different microcavity decay rates. The points below each line correspond to the condensate region. The red line corresponds to the case of infinite microcavity lifetime, where condensation occurs at the onset of inversion. The parameter region of the condensed phase is smaller for larger radiative decay, as it implies more inversion is required to produce condensation. }
     \label{fig:phasediagram}
\end{figure}

Eq. (\ref{eq:sscond}) and the equations-of-motion can be solved numerically to determine the condensate occupation $N_c$ in the steady-state. This solution can be used to map the phase diagram, i.e., the region of parameter space in which $N_c\neq 0$. Fig.\ \ref{fig:phasediagram} shows a numerically computed phase diagram for condensation, in terms of the temperatures of the hot (pump) and cold (lattice) heat baths, for several values of the decay rate $\gamma$. The condensed state, with $N_c\neq 0$, lies below each curve, with the normal state, $N_c=0$, above it. For these parameters the transition is continuous. It can be seen that the phase boundaries approach the red diagonal line as $\gamma\rightarrow 0$. This corresponds to the ideal thermodynamically-reversible limit~\cite{scovil_three-level_1959}, in which condensation occurs when inversion is reached, $\rho_{33}/\rho_{22}=1$.  As noted by Scovil and Schulz-DuBois~\cite{scovil_three-level_1959}, the requirement of inversion implies laser action is bounded by the Carnot efficiency, and reaches it in the reversible limit where inversion first appears. Generalizing their argument to include the chemical potential of the hot reservoir, the condition for inversion \begin{equation} \frac{\rho_{33}}{\rho_{22}}=\frac{\rho_{33}}{\rho_{11}}\frac{\rho_{11}}{\rho_{22}}=e^{-\beta_h (e_3-\mu)}e^{\beta_c e_2}>1,\end{equation} implies that the efficiency \begin{align}\eta=(W_{out}-W_{in})/Q_{in} =(\omega-\mu)/(e_3-\mu),\label{eq:standardeff}\end{align} is less than the Carnot efficiency $\eta_C=1-T_c/T_h$. Here we have used  $e_3-e_2=\omega$ ($\Delta =0$) and  identified $\omega$ as the work output per cycle, $\mu$ as the work input, and $e_3-\mu$ as the heat input (with $e_1=0$).  

The continuous transition can be understood using a perturbative expansion of the steady-state in $\Omega$. In the normal state we have $\Omega=0$ so $\theta\rightarrow 0$, $\Tilde{e}_2=e_2, \Tilde{e}_3=e_3-\omega$. The equations-of-motion reduce to the standard Lamb equations for a three-level laser~\cite{geva_quantum_1996}, given in the Appendix. The steady-state has the populations $\rho_{33}/\rho_{11}$ and $\rho_{22}/\rho_{11}$ in equilibrium with the hot and cold baths, respectively, and vanishing coherences. Expanding around this solution gives the steady-state  coherence to first order in $\Omega$, and hence, using Eq. (\ref{eq:gain}), the scattering rate \begin{equation} R=\frac{\Omega^2\left(\rho_{33}-\rho_{22}\right)} {\gamma^c_\downarrow+\gamma^h_\downarrow}= \frac{4g_c^2 N\left(\rho_{33}-\rho_{22}\right)} {\gamma^c_\downarrow+\gamma^h_\downarrow}.\label{eq:lineargain}\end{equation} Here $\gamma^c_\downarrow$ ($\gamma^h_\downarrow$) are the emission rates into the cold (hot) bath, respectively. Eq. (\ref{eq:lineargain}) is the Fermi golden rule expression for scattering into a final state, with a linewidth generated by the emission into the baths. It defines, via Eq. (\ref{eq:sscond}), the  critical inversion at which gain exceeds loss, and hence the phase boundary. The energies involved in the cycle are unaffected by the condensate to this order, so the efficiency, Eq. (\ref{eq:standardeff}), is constant.

The red line in Fig.\ \ref{fig:phasediagram} separates the diagram into two regions: above it, the efficiency set by the energies is greater than the Carnot efficiency, and condensation can never occur. It emerges immediately below this line for $\gamma\rightarrow 0$, where condensation is supported by an infinitesimal power flow from the working medium. In this limit the threshold for condensation corresponds to the condition that the entropy changes of the hot and cold reservoirs balance, \begin{equation} \frac{e_2}{T_c}=\frac{e_3-\mu}{T_h}. \end{equation} For non-zero $\gamma$, condensation requires finite, and therefore irreversible, power flows from the working medium, which will produce a lower-than-Carnot efficiency for given bath temperatures. The phase boundary therefore departs from the reversible line; the non-zero $\gamma$ implies that the temperature difference $T_h-T_c$ required to drive condensation is increased, as it must overcome the loss. In more practical terms, a larger $\gamma$ requires a larger inversion to overcome the loss, and hence a higher (lower) temperature for the hot (cold) bath. 


\begin{figure}[t]
    \centering
     \includegraphics[width=1\linewidth]{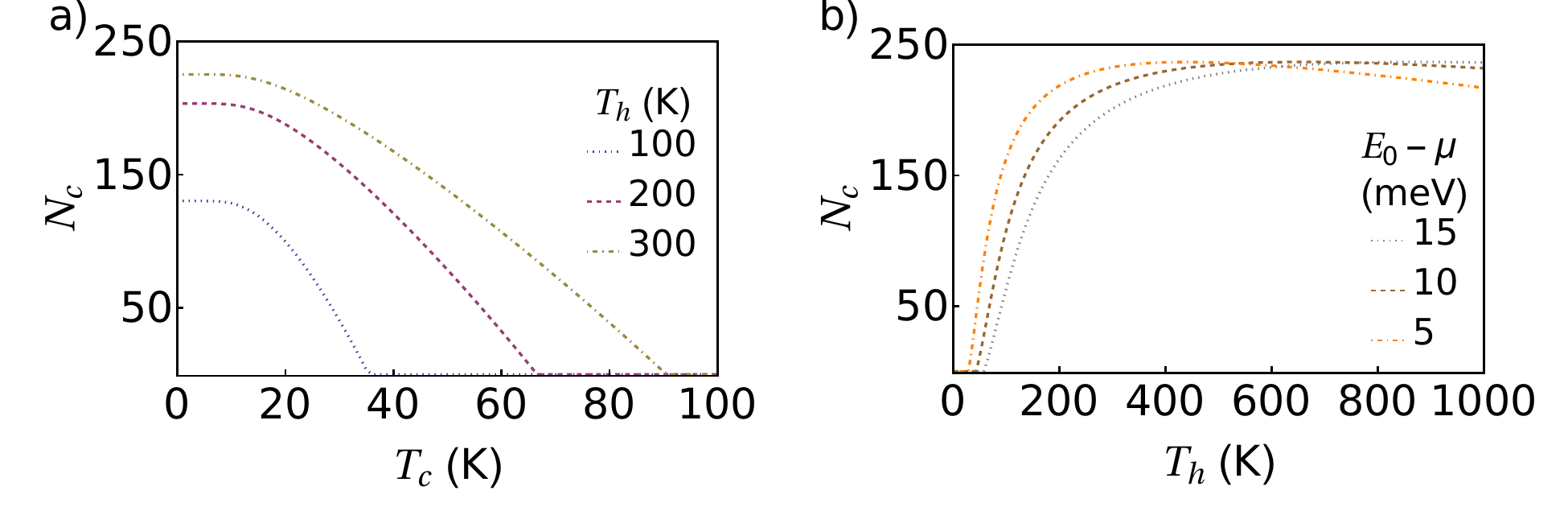}
    \caption{Size of the condensate as a function of the temperature of the cold (a) and hot (b) baths for various temperatures of the hot bath, and chemical potentials, respectively. A second order phase transition is observed from the normal state with $N_c =0$ to a condensed state with non-vanishing occupation.}
     \label{fig:BECsize}
\end{figure}

To determine the steady-state occupation in the condensed state we solve Eq. (\ref{eq:sscond}) using the steady-state calculated numerically. This incorporates the dependence of the dressed-state energies and wavefunctions on the condensate occupation, which provides the nonlinearities, beyond Eq. (\ref{eq:lineargain}), that are needed to stabilize the condensate at a finite density. Results are shown in Fig.\ \ref{fig:BECsize}, as functions of the two bath temperatures. 

Considering first the dependence on $T_h$, we see that the condensate size increases rapidly with $T_h$ once the threshold is crossed, and then saturates or decreases. This is expected from the occupation of the hot bath at the energy of the upper level ($\approx e_3$). If we consider the equilibrium of the upper level and the hot bath only, its population will be $(1+e^{-\beta_h(e_3-\mu)})^{-1}$, which is negligible until $k T_h\sim (e_3-\mu)$, at which point it rapidly grows before saturating at $0.5$. The small decrease could be explained by noting that $\gamma_h^\downarrow$, approximately proportional to $(1+n_h(e_3))$, increases with temperature due to the stimulated emission into the hot reservoir. This will broaden the line and hence, as expected from Eq. (\ref{eq:lineargain}), reduce the gain. Such a mechanism has been shown to produce an upper critical threshold, i.e., a maximum $T_h$, in the three-level laser model\ \cite{li_quantum_2017}.

The dependence of the condensate size on $T_c$ can be understood analogously in terms of the thermal occupation of the cold bath, i.e., the phonons. Below the critical temperature the condensate occupation increases smoothly, and plateaus at very low temperatures. The critical $T_c$ arises here from the requirements that the lower state of the $3-2$ transition should be depopulated sufficiently to generate inversion -- in other words, the phonons must cool the high-momentum excitons, here represented by the pump and the idler states, to a low temperature so that the idler population is small. The saturation at low temperatures arises from the assumption of a single energy gap for the phonon relaxation, which implies that the equilibrium idler population is effectively zero for temperatures $kT_c\ll e_2$.


\begin{figure}[t]
    \centering
     \includegraphics[width=.9\linewidth]{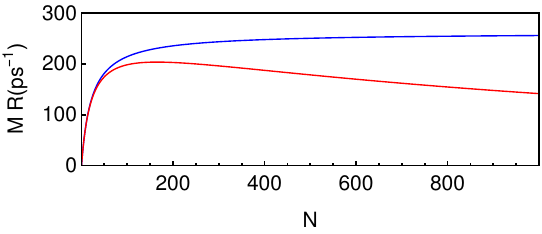}
    \caption{Total in-scattering rate into the condensate as a function of its size, with a Lorentzian (red) and constant (blue)  phonon spectral density. The rate reaches a maximum value for both spectral densities, however the Lorentzian shape of the spectral density causes a reduction in $R$ for large condensates.}
     \label{fig:gaincurve}
\end{figure}

Fig.\ \ref{fig:gaincurve} shows the total scattering rate into the condensate, $MR(N)$, in the steady-state, as function of the condensate size $N$. This shows the existence of a maximum power output of the three-level system. Such maxima are general features of the three-level heat engine~\cite{geva_quantum_1996}, caused by the energy-dependence of the scattering rates, which are sampled at the dressed-state energies $\Tilde{e}_{2,3}$ with splitting $\sim \Omega$ when $\Omega \gg \Delta$. In Ref.~\onlinecite{geva_quantum_1996} the existence of a maximum power was attributed to the frequency-dependence of the hot bath occupation function. Although this thermal factor will produce such an effect in general, it cannot explain the particular peak in Fig.\ \ref{fig:gaincurve}, since here $k T_h$ is much larger than the energy shifts at the maximum power point. It arises, instead, from the spectral function of the cold (phonon) environment, which we have modeled as a Lorentzian of width $g_{ph}$. When $N=0$ the dressed-states energies lie at the peak of this Lorentzian, but they move away as $\Omega\sim\sqrt{N}$ increases, reducing the emission rate of phonons into the environment and slowing the cycle. This interpretation is consistent with the result for a flat phonon spectral density, shown for comparison. Although the assumption of a Lorentzian spectral density for the phonons is not necessarily realistic, the heat baths relevant for polariton condensates do have spectral structure. This can be expected to lead to a maximum power once the condensate nonlinearities become comparable to either the scale of the structure or the temperatures.

\begin{figure}[t]
    \centering
     \includegraphics[width=.9\linewidth]{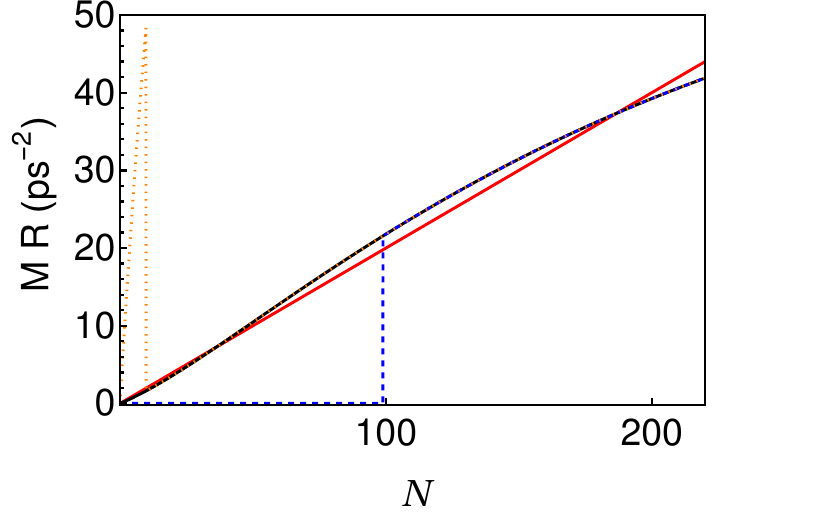}
    \caption{In-scattering rates with $\Delta=0.3$ meV and $E_0-e_3$=0 (black solid), 0.33 meV (blue dotted), -0.2 meV (orange dotted). The red curve is the loss rate. These parameters lead to in-scattering rates with a concave region (black solid and orange dotted) and discontinuities (blue dotted and orange dotted). Such forms give rise to first-order transitions and bistability (see text).}
     \label{fig:1storder}
\end{figure}   

In some cases there are additional phenomena which can arise from the non-trivial spectral densities of the baths. Fig.\ \ref{fig:1storder} shows the gain as a function of condensate size for three different sets of parameters, revealing two unexpected effects. The first is the presence of discontinuities in the steady-state, either connecting a region of zero gain to non-zero gain, or causing a sudden drop in the gain. These arise from the interplay between the energy shifts of the driving and the spectral density of the hot bath, which we have assumed to have a step-function onset at energy $E_0$. The dashed-blue curve arises in a situation where the upper level of the heat engine lies just below the hot bath, $e_3<E_0$, so that there is no population of the upper level in the absence of the condensate, and hence no gain. However, as the condensate size increases the upper dressed state, $\Tilde{e}_3$, moves to higher energies, and crosses $E_0$, at which point gain appears. A related situation, shown by the orange dotted curve, appears if $e_3$ starts just above $E_0$. In that case, while the two dressed states start in the band, as the condensate occupation increases the lower-energy one drops below $E_0$, and the gain suddenly decreases. The second, more subtle, effect is also visible in this curve, as well as the black curve, which corresponds to the case where $e_3$ starts exactly at $E_0$. This is the presence of a concave part of the gain curve at small $N$. We suggest this is because there is also a small positive detuning here, so that the upper dressed state is composed mostly of $|3\rangle$ in the limit $N\rightarrow 0$. However, as the splitting $\Omega$ reaches the detuning $\Delta$, this component of the upper dressed state -- which is the only one pumped by the hot bath at this point in the curve -- begins to reduce significantly, suppressing the gain. We note this concave form of $R(N)$ implies the transition is first-order, with the crossing point solving $MR(N)=\gamma N$ first appearing at a non-zero $N$. A first-order behavior can also be expected from the physics of the step-like curve which, for the more realistic case of a smoothly-increasing spectral density, will become a smooth but concave $R(N)$. In addition, we expect these forms to lead to bistability of the condensed states, since they have two intersections with the loss curve.

The physics of these effects lead us to suggest they could be achieved experimentally by constructing a system in which the source of the pairwise scattering, i.e. the bottleneck state, lies in a region where either the population or spectral density of the exciton reservoir increases with increasing energy. A natural way to do this would be to use resonant pumping at an energy above the pairwise resonance. We note that, although we have assumed a thermal population, the expressions for the rates, Eqs. (\ref{eq:coldcurbare}) and (\ref{eq:hotcurbare}) or (\ref{eq:coldcur}) and (\ref{eq:hotcur}), can be used also in the non-thermal case. 

\begin{figure}[t]
    \centering
     \includegraphics[width=1\linewidth]{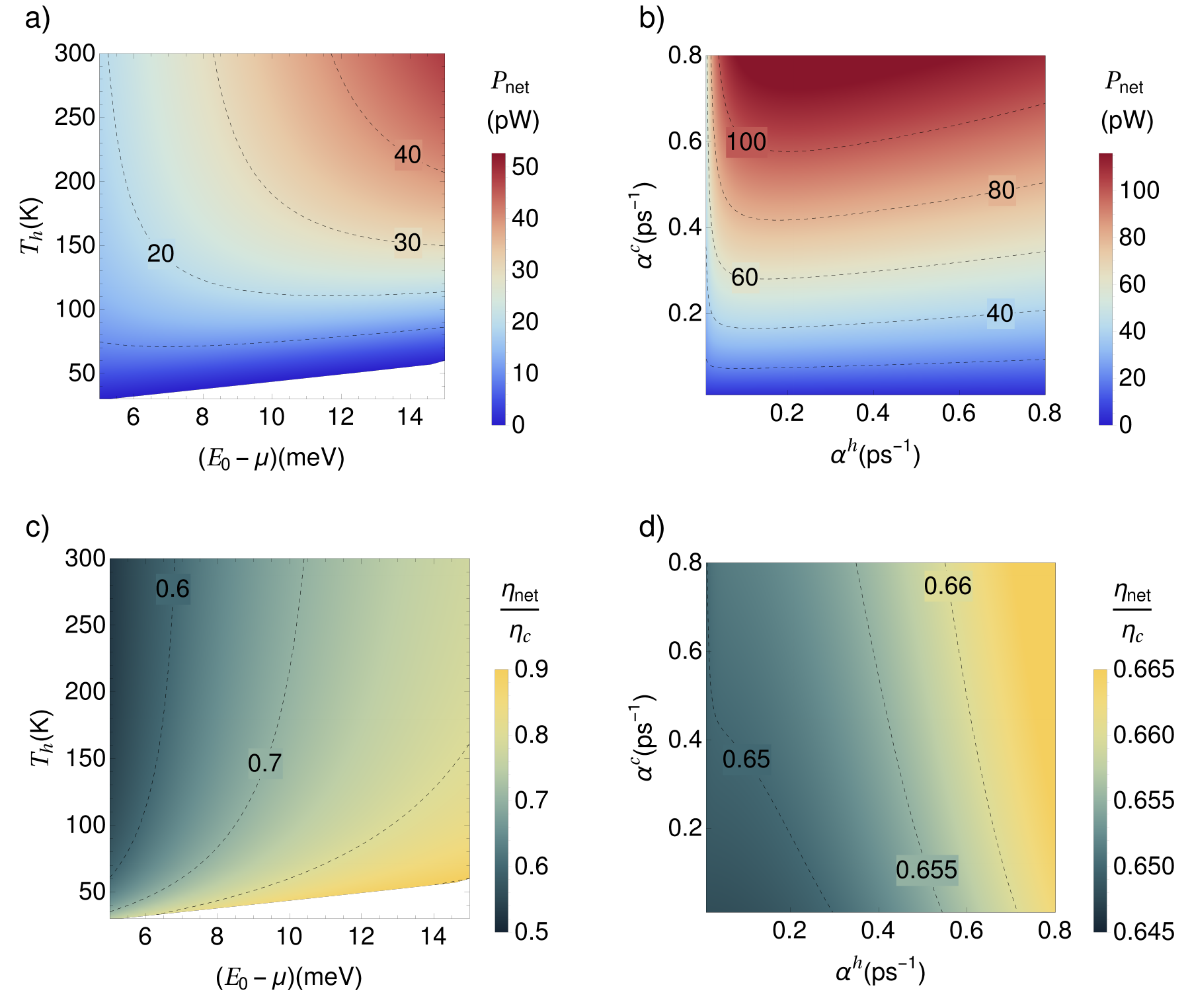}
    \caption{Net output power (top row) and efficiency (bottom row) of a steady-state condensate, as functions of the temperature and chemical potential of the hot bath (left column), and of the coupling strengths to the two baths (right column). The chemical potential is measured relative to the lowest state of the pump reservoir, $E_0$, and the efficiencies are normalized by the Carnot efficiency. The net output power grows if $T_h$ or $(E_0 - \mu)$ increase and has stronger dependence on the coupling strength with the cold than with the hot bath. The efficiency shows only a weak dependence on the parameters considered.}
     \label{fig:thermo2}
\end{figure}

Figure~\ref{fig:thermo2} shows some results for the power output and efficiency of the condensate. The white regions denote the non-condensed phase, and the colors depict the net power $P_\text{out} - P_\text{in}$, and net efficiency $- (P_\text{out} - P_\text{in})/Q_\text{h}$, in the steady-state. The two plots in the left column show the effects of varying the hot bath temperature and chemical potential, while the right column shows the corresponding effects of varying the coupling strengths. The net power increases with increasing hot bath temperature, in line with the condensate occupation shown in Fig.\ \ref{fig:BECsize}. However, while the condensate occupation increases with increasing chemical potential, the net power output decreases. This is because although increasing $\mu$ increases the cycle rate and hence the gross power, $\omega R=\omega \gamma N/M$, it also increases the work input, giving an overall decrease in the net power, $(\omega-\mu)R$. Considered as functions of the coupling strengths, the net power output increases significantly with an increasing coupling to the cold bath. However, the power output depends only weakly on the hot bath coupling parameter, and can even decrease as it increases. We suggest that the weak dependence arises because, for the temperatures chosen, the cycle rate is limited by emission into the cold bath rather than absorption from the hot one. This emphasizes the need for rapid phonon thermalization in the high-momentum excitons, in order to drive condensation (see Fig.\ \ref{fig:diagram}). 

The efficiency, as a function of temperature and coupling constants, is shown in the lower two panels of Fig.~\ref{fig:thermo2}. It is shown relative to the Carnot efficiency, which explains most of the dependence on the temperatures; the energy shifts are small for these parameters, so that $\eta$ is well approximated by Eq. (\ref{eq:standardeff}). The efficiency at threshold is below the Carnot efficiency due to the finite loss rate. Since the efficiency is determined almost entirely by the energy levels, there is only a very small effect of the coupling strengths. 

So far, we have considered condensation in the regime $\omega>\mu$. Here the work output per cycle, $\omega$, exceeds the work input, $\mu$, so that condensation is only possible with the conversion of heat to work. The condensate in this case operates as a heat engine, and requires a higher temperature for the exciton reservoir than for the phonons. However, if $\omega<\mu$ the work output per cycle, $\omega$, is less than the work input per cycle, $\mu$, so the machine operates not as a heat engine, but as a dissipator or refrigerator. In these modes heat is not converted to work, as in a heat engine, but rather the excess work, $\mu-\omega$, is dissipated as heat.

\begin{figure}[t]
    \centering
     \includegraphics[width=1\linewidth]{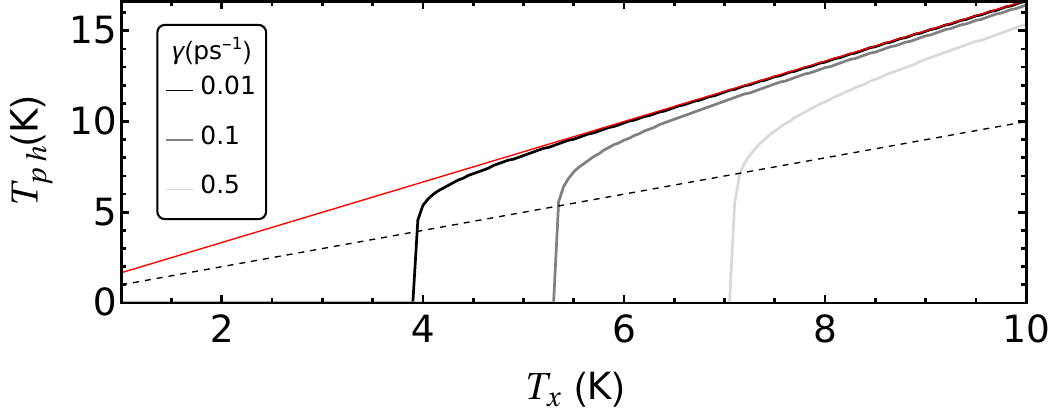}
    \caption{Phase diagram of condensation for $\omega<\mu = 1.002$ eV, as a function of the temperatures of the exciton $T_x$ and phonon $T_{ph}$ reservoir, for three different values of decay rates. The area below each curve corresponds to the parameter region where condensation can occur. The red line corresponds to the reversible limit of the thermal machine, and indicates the phase boundary when $\gamma \rightarrow 0$. }
     \label{fig:refrigeratorex}
\end{figure}

Fig.~\ref{fig:refrigeratorex} shows the phase diagram of condensation when $\omega - \mu = 2$ meV. To avoid confusion we here use $T_x$ and $T_{ph}$ to refer to the temperatures of the exciton and phonon reservoir, instead of the subscripts denoting hot and cold. The condensate occurs in the region below and to the right of the curves, and it can be seen that at the onset of condensation $T_{ph}>T_x$. The excess work here is deposited as heat in the higher-temperature reservoir, corresponding to a flow of heat against the temperature gradient. The dashed line corresponds to  $T_x =T_{ph}$, so that in the region below this the excess work is dissipated in the lower-temperature reservoir.

\section{Conclusions}

In summary, we have argued that a driven-dissipative condensate is a form of heat engine, and as such requires consideration of both a hot and cold bath. We considered in depth the case of a microcavity polariton condensate, and constructed a minimal heat-engine model that captures the key processes involved in condensation. Our results show how condensation is determined by the temperatures of both the hot (exciton) and cold (phonon) bath, and that the temperature difference between these must exceed that required in the reversible limit in order to overcome the polariton loss. They also emphasize the importance of rapid cooling in the high-momentum exciton states, in order to maintain effective scattering into the condensate. The maximum power output of the condensate shows effects of the spectral densities of the environment, and the formation of dressed states, which can also produce unusual phenomena including a first-order phase boundary and bistability.

Our results provide guidance for extending the regimes and systems which support condensation, and our methods could be extended to consider other examples such as the condensation~\cite{klaers_boseeinstein_2010} and thermalization~\cite{busley_sunlight-pumped_2023} of photons. Furthermore, it would also be interesting to explore the reversed operation of the heat-engine and the possibilities for phonon refrigeration~\cite{klembt_exciton-polariton_2015}.

One extension of our work would be to include a distribution of energies for the states in the working medium, i.e. inhomogeneous broadening. In this case the gain, given by Eq.~\ref{eq:lineargain} and appearing in the threshold Eq.~\eqref{eq:sscond}, would be replaced its average over the broadened line. While this would modify the details of the phase boundary, and make condensation more difficult due to the dispersion of the gain over different frequencies, we would not expect it to have dramatic effects unless the broadening becomes comparable to the scale of the dispersion relation. Small broadenings could, however, modify some of the details of the results in the nonlinear regime, where they would be relevant in comparison to the energy shifts $\Omega$ or features in the bath spectral densities. We have found such effects in preliminary work; for example, the peak in Fig.~\ref{fig:gaincurve} is replaced, over the range shown, by a plateau, if the levels $e_2$ and $e_3$ are broadened by $2\ \mathrm{meV}$ while keeping $e_3-e_2$ fixed.

\begin{acknowledgments}
P.R.E. acknowledges support from Science Foundation Ireland (21-FFP-10142). 
\end{acknowledgments}

\section*{Authors declarations}
\subsection*{Conflict of Interest}
The authors have no conflicts to disclose. 

\subsection*{Author Contributions}
{\bf Luisa Toledo Tude:} Conceptualization (supporting), formal analysis (lead), methodology (lead), software (lead), visualization (lead), writing -- original draft preparation (lead), writing -- review and editing (equal). {\bf Paul Eastham:} Conceptualization (lead), supervision (lead), funding acquisition (lead), methodology (supporting), supervision (lead), writing -- original draft preparation (supporting), writing -- review and editing (equal). 

\section*{Data availability}

The code used to generate the data that support the findings of this study is openly available in Zenodo at https://doi.org/10.5281/zenodo.10822700. In order to meet institutional and research funder open access requirements, any
accepted manuscript arising shall be open access under a Creative Commons Attribution (CC BY) reuse license with zero embargo.


\appendix

\section{Equations-of-motion}

The equations-of-motion for the diagonal elements of the density matrix, in the bare basis, are given in Eqs. (\ref{eq:dmeom11}--\ref{eq:dmeom33}). We give here the remaining terms, considering the contributions from the two baths separately. We use an abbreviated notation in which subscripts denote the frequencies at which the spectral densities and bath occupations are sampled, so that in expressions for the cold (hot) bath we have $J_i=J_c(\Tilde{e}_i), n_i=n_c(\Tilde{e}_i)$ ($J_i=J_h(\Tilde{e}_i+\omega), n_i=n_h(\Tilde{e}_i+\omega)$). For the cold bath we have
\begin{widetext}
\begin{align}
    \dot\rho_{23}|_c=-\frac{\pi}{2} & \{ [\rho_{11} (J_2 n_2-J_3 n_3) +\rho_{33} (J_3 (1+n_3) -J_2 
    (1+n_2))]\sin\theta \nonumber \\ & + \rho_{23} [J_2 (1+n_2) (1+\cos\theta ) + J_3 (1+n_3)(1-\cos\theta ) ]\},
\label{eq:odeom23c}\end{align}
\begin{align}
    \dot{\rho}_{12}|_c=-\frac{\pi}{2}\{( & \rho_{12}-\rho_{21})[J_2(1+2n_2)(1+\cos\theta) + J_3(1+2n_3)(1-\cos\theta)] \nonumber \\ 
    + &( \rho_{13}-\rho_{31})[J_3(1+n_3)-J_2(1+n_2)]\sin\theta
    \},
\end{align}
\begin{align}
    \dot\rho_{13}|_c=-\frac{\pi}{2}&\{\rho_{13}[J_2n_2(1+\cos\theta)+J_3n_3(1-\cos\theta)] \nonumber \\ & + \rho_{21}[J_2n_2-J_3n_3]\sin\theta\}.
\end{align}

The corresponding expressions for the hot bath are 
\begin{align}
    \dot\rho_{23}|_h=-\frac{\pi}{2} & \{ [\rho_{11} (J_2 n_2-J_3 n_3) +\rho_{22} (J_3 (1+n_3) -J_2 
    (1+n_2))]\sin\theta \nonumber \\ & + \rho_{23} [J_2 (1+n_2) (1-\cos\theta ) + J_3 (1+n_3)(1+\cos\theta ) ]\},
\end{align}
\begin{align}
    \dot{\rho}_{13}|_h=-\frac{\pi}{2}\{\rho_{13}[J_3(1+2n_3)(1+\cos\theta) + J_2(1+2n_2)(1-\cos\theta)] \nonumber \\ 
    +  \rho_{12}[J_3(1+n_3)-J_2(1+n_2)]\sin\theta
    \},
\end{align}
\begin{align}
    \dot\rho_{12}|_h=-\frac{\pi}{2}&\{\rho_{12}[J_3n_3(1+\cos\theta)+J_2n_2(1-\cos\theta)]\}.\label{eq:odeom12h}
\end{align}
    
\end{widetext}
In the limit $\theta\rightarrow 0$, the energies of the dressed states reduce to the original energies, $\Tilde{e}_2=e_2, \Tilde{e}_3=e_3-\omega$, and the equations-of-motion become the standard Lamb equations for a three-level laser~\cite{geva_quantum_1996}. The population transfer rates become \begin{align}
R^c&=2\pi J_2 (1+n_2)\rho_{22}-2\pi J_2 n_2\rho_{11} \nonumber \\&=\gamma^c_\downarrow\rho_{22}-\gamma^c_\uparrow\rho_{11}, \label{eq:lambcold}\\
-R^h&=2\pi J_3 (1+n_3)\rho_{33}-2\pi J_3 n_3\rho_{11}\nonumber \\&=\gamma^h_\downarrow\rho_{33}-\gamma^h_\uparrow\rho_{11}. \label{eq:lambhot}
 \end{align} The dissipative contributions to the equations-of-motion for the off-diagonal elements of the density matrix become\begin{align}
    \dot\rho_{23}=-\frac{1}{2}(\gamma^c_\downarrow+\gamma^h_\downarrow)\rho_{23}, \label{eq:lambcoh1} \\
    \dot{\rho}_{13}=-\frac{1}{2}(\gamma^h_\uparrow+\gamma^h_\downarrow+\gamma_c^\uparrow)\rho_{13}, \label{eq:lambcoh2}\\
    \dot\rho_{12}=-\frac{1}{2}(\gamma^c_\uparrow+\gamma^c_\downarrow+\gamma_h^\uparrow)\rho_{12}. \label{eq:lambcoh3}
\end{align} where a counterrotating term from the cold bath has been neglected in Eq. (\ref{eq:lambcoh3}). These expressions describe the decay of the coherences, with the rates related in the expected way to those of the populations.

\section*{References}
\bibliography{polaritonthermo}

\begin{thebibliography}{39}%
\makeatletter
\providecommand \@ifxundefined [1]{%
 \@ifx{#1\undefined}
}%
\providecommand \@ifnum [1]{%
 \ifnum #1\expandafter \@firstoftwo
 \else \expandafter \@secondoftwo
 \fi
}%
\providecommand \@ifx [1]{%
 \ifx #1\expandafter \@firstoftwo
 \else \expandafter \@secondoftwo
 \fi
}%
\providecommand \natexlab [1]{#1}%
\providecommand \enquote  [1]{``#1''}%
\providecommand \bibnamefont  [1]{#1}%
\providecommand \bibfnamefont [1]{#1}%
\providecommand \citenamefont [1]{#1}%
\providecommand \href@noop [0]{\@secondoftwo}%
\providecommand \href [0]{\begingroup \@sanitize@url \@href}%
\providecommand \@href[1]{\@@startlink{#1}\@@href}%
\providecommand \@@href[1]{\endgroup#1\@@endlink}%
\providecommand \@sanitize@url [0]{\catcode `\\12\catcode `\$12\catcode
  `\&12\catcode `\#12\catcode `\^12\catcode `\_12\catcode `\%12\relax}%
\providecommand \@@startlink[1]{}%
\providecommand \@@endlink[0]{}%
\providecommand \url  [0]{\begingroup\@sanitize@url \@url }%
\providecommand \@url [1]{\endgroup\@href {#1}{\urlprefix }}%
\providecommand \urlprefix  [0]{URL }%
\providecommand \Eprint [0]{\href }%
\providecommand \doibase [0]{https://doi.org/}%
\providecommand \selectlanguage [0]{\@gobble}%
\providecommand \bibinfo  [0]{\@secondoftwo}%
\providecommand \bibfield  [0]{\@secondoftwo}%
\providecommand \translation [1]{[#1]}%
\providecommand \BibitemOpen [0]{}%
\providecommand \bibitemStop [0]{}%
\providecommand \bibitemNoStop [0]{.\EOS\space}%
\providecommand \EOS [0]{\spacefactor3000\relax}%
\providecommand \BibitemShut  [1]{\csname bibitem#1\endcsname}%
\let\auto@bib@innerbib\@empty
\bibitem [{\citenamefont {Keeling}\ and\ \citenamefont
  {{K{\'e}na-Cohen}}(2020)}]{keeling_boseeinstein_2020}%
  \BibitemOpen
  \bibfield  {author} {\bibinfo {author} {\bibfnamefont {J.}~\bibnamefont
  {Keeling}}\ and\ \bibinfo {author} {\bibfnamefont {S.}~\bibnamefont
  {{K{\'e}na-Cohen}}},\ }\bibfield  {title} {\enquote {\bibinfo {title}
  {Bose--{{Einstein Condensation}} of {{Exciton-Polaritons}} in {{Organic
  Microcavities}}},}\ }\href
  {https://doi.org/10.1146/annurev-physchem-010920-102509} {\bibfield
  {journal} {\bibinfo  {journal} {Annu. Rev. Phys. Chem.}\ }\textbf {\bibinfo
  {volume} {71}},\ \bibinfo {pages} {435--459} (\bibinfo {year}
  {2020})}\BibitemShut {NoStop}%
\bibitem [{\citenamefont {Deng}, \citenamefont {Haug},\ and\ \citenamefont
  {Yamamoto}(2010)}]{deng_exciton-polariton_2010}%
  \BibitemOpen
  \bibfield  {author} {\bibinfo {author} {\bibfnamefont {H.}~\bibnamefont
  {Deng}}, \bibinfo {author} {\bibfnamefont {H.}~\bibnamefont {Haug}},\ and\
  \bibinfo {author} {\bibfnamefont {Y.}~\bibnamefont {Yamamoto}},\ }\bibfield
  {title} {\enquote {\bibinfo {title} {Exciton-polariton {{Bose-Einstein}}
  condensation},}\ }\href {https://doi.org/10.1103/RevModPhys.82.1489}
  {\bibfield  {journal} {\bibinfo  {journal} {Rev. Mod. Phys.}\ }\textbf
  {\bibinfo {volume} {82}},\ \bibinfo {pages} {1489--1537} (\bibinfo {year}
  {2010})}\BibitemShut {NoStop}%
\bibitem [{\citenamefont {Klaers}\ \emph {et~al.}(2010)\citenamefont {Klaers},
  \citenamefont {Schmitt}, \citenamefont {Vewinger},\ and\ \citenamefont
  {Weitz}}]{klaers_boseeinstein_2010}%
  \BibitemOpen
  \bibfield  {author} {\bibinfo {author} {\bibfnamefont {J.}~\bibnamefont
  {Klaers}}, \bibinfo {author} {\bibfnamefont {J.}~\bibnamefont {Schmitt}},
  \bibinfo {author} {\bibfnamefont {F.}~\bibnamefont {Vewinger}},\ and\
  \bibinfo {author} {\bibfnamefont {M.}~\bibnamefont {Weitz}},\ }\bibfield
  {title} {\enquote {\bibinfo {title} {Bose--{{Einstein}} condensation of
  photons in an optical microcavity},}\ }\href
  {https://doi.org/10.1038/nature09567} {\bibfield  {journal} {\bibinfo
  {journal} {Nature}\ }\textbf {\bibinfo {volume} {468}},\ \bibinfo {pages}
  {545--548} (\bibinfo {year} {2010})}\BibitemShut {NoStop}%
\bibitem [{\citenamefont {Kirton}\ and\ \citenamefont
  {Keeling}(2013)}]{kirton_nonequilibrium_2013}%
  \BibitemOpen
  \bibfield  {author} {\bibinfo {author} {\bibfnamefont {P.}~\bibnamefont
  {Kirton}}\ and\ \bibinfo {author} {\bibfnamefont {J.}~\bibnamefont
  {Keeling}},\ }\bibfield  {title} {\enquote {\bibinfo {title} {Nonequilibrium
  {{Model}} of {{Photon Condensation}}},}\ }\href
  {https://doi.org/10.1103/PhysRevLett.111.100404} {\bibfield  {journal}
  {\bibinfo  {journal} {Phys. Rev. Lett.}\ }\textbf {\bibinfo {volume} {111}},\
  \bibinfo {pages} {100404} (\bibinfo {year} {2013})}\BibitemShut {NoStop}%
\bibitem [{\citenamefont {Hakala}\ \emph {et~al.}(2018)\citenamefont {Hakala},
  \citenamefont {Moilanen}, \citenamefont {V{\"a}kev{\"a}inen}, \citenamefont
  {Guo}, \citenamefont {Martikainen}, \citenamefont {Daskalakis}, \citenamefont
  {Rekola}, \citenamefont {Julku},\ and\ \citenamefont
  {T{\"o}rm{\"a}}}]{hakala_boseeinstein_2018}%
  \BibitemOpen
  \bibfield  {author} {\bibinfo {author} {\bibfnamefont {T.~K.}\ \bibnamefont
  {Hakala}}, \bibinfo {author} {\bibfnamefont {A.~J.}\ \bibnamefont
  {Moilanen}}, \bibinfo {author} {\bibfnamefont {A.~I.}\ \bibnamefont
  {V{\"a}kev{\"a}inen}}, \bibinfo {author} {\bibfnamefont {R.}~\bibnamefont
  {Guo}}, \bibinfo {author} {\bibfnamefont {J.-P.}\ \bibnamefont
  {Martikainen}}, \bibinfo {author} {\bibfnamefont {K.~S.}\ \bibnamefont
  {Daskalakis}}, \bibinfo {author} {\bibfnamefont {H.~T.}\ \bibnamefont
  {Rekola}}, \bibinfo {author} {\bibfnamefont {A.}~\bibnamefont {Julku}},\ and\
  \bibinfo {author} {\bibfnamefont {P.}~\bibnamefont {T{\"o}rm{\"a}}},\
  }\bibfield  {title} {\enquote {\bibinfo {title} {Bose--{{Einstein}}
  condensation in a plasmonic lattice},}\ }\href
  {https://doi.org/10.1038/s41567-018-0109-9} {\bibfield  {journal} {\bibinfo
  {journal} {Nature Phys}\ }\textbf {\bibinfo {volume} {14}},\ \bibinfo {pages}
  {739--744} (\bibinfo {year} {2018})}\BibitemShut {NoStop}%
\bibitem [{\citenamefont {Demokritov}\ \emph {et~al.}(2006)\citenamefont
  {Demokritov}, \citenamefont {Demidov}, \citenamefont {Dzyapko}, \citenamefont
  {Melkov}, \citenamefont {Serga}, \citenamefont {Hillebrands},\ and\
  \citenamefont {Slavin}}]{demokritov_boseeinstein_2006}%
  \BibitemOpen
  \bibfield  {author} {\bibinfo {author} {\bibfnamefont {S.~O.}\ \bibnamefont
  {Demokritov}}, \bibinfo {author} {\bibfnamefont {V.~E.}\ \bibnamefont
  {Demidov}}, \bibinfo {author} {\bibfnamefont {O.}~\bibnamefont {Dzyapko}},
  \bibinfo {author} {\bibfnamefont {G.~A.}\ \bibnamefont {Melkov}}, \bibinfo
  {author} {\bibfnamefont {A.~A.}\ \bibnamefont {Serga}}, \bibinfo {author}
  {\bibfnamefont {B.}~\bibnamefont {Hillebrands}},\ and\ \bibinfo {author}
  {\bibfnamefont {A.~N.}\ \bibnamefont {Slavin}},\ }\bibfield  {title}
  {\enquote {\bibinfo {title} {Bose--{{Einstein}} condensation of
  quasi-equilibrium magnons at room temperature under pumping},}\ }\href
  {https://doi.org/10.1038/nature05117} {\bibfield  {journal} {\bibinfo
  {journal} {Nature}\ }\textbf {\bibinfo {volume} {443}},\ \bibinfo {pages}
  {430--433} (\bibinfo {year} {2006})}\BibitemShut {NoStop}%
\bibitem [{\citenamefont {Scovil}\ and\ \citenamefont
  {{Schulz-DuBois}}(1959)}]{scovil_three-level_1959}%
  \BibitemOpen
  \bibfield  {author} {\bibinfo {author} {\bibfnamefont {H.~E.~D.}\
  \bibnamefont {Scovil}}\ and\ \bibinfo {author} {\bibfnamefont {E.~O.}\
  \bibnamefont {{Schulz-DuBois}}},\ }\bibfield  {title} {\enquote {\bibinfo
  {title} {Three-{{Level Masers}} as {{Heat Engines}}},}\ }\href
  {https://doi.org/10.1103/PhysRevLett.2.262} {\bibfield  {journal} {\bibinfo
  {journal} {Phys. Rev. Lett.}\ }\textbf {\bibinfo {volume} {2}},\ \bibinfo
  {pages} {262--263} (\bibinfo {year} {1959})}\BibitemShut {NoStop}%
\bibitem [{\citenamefont {Geusic}, \citenamefont {{Schulz-DuBios}},\ and\
  \citenamefont {Scovil}(1967)}]{geusic_quantum_1967}%
  \BibitemOpen
  \bibfield  {author} {\bibinfo {author} {\bibfnamefont {J.~E.}\ \bibnamefont
  {Geusic}}, \bibinfo {author} {\bibfnamefont {E.~O.}\ \bibnamefont
  {{Schulz-DuBios}}},\ and\ \bibinfo {author} {\bibfnamefont {H.~E.~D.}\
  \bibnamefont {Scovil}},\ }\bibfield  {title} {\enquote {\bibinfo {title}
  {Quantum {{Equivalent}} of the {{Carnot Cycle}}},}\ }\href
  {https://doi.org/10.1103/PhysRev.156.343} {\bibfield  {journal} {\bibinfo
  {journal} {Phys. Rev.}\ }\textbf {\bibinfo {volume} {156}},\ \bibinfo {pages}
  {343--351} (\bibinfo {year} {1967})}\BibitemShut {NoStop}%
\bibitem [{\citenamefont {Kosloff}(1984)}]{kosloff_quantum_1984}%
  \BibitemOpen
  \bibfield  {author} {\bibinfo {author} {\bibfnamefont {R.}~\bibnamefont
  {Kosloff}},\ }\bibfield  {title} {\enquote {\bibinfo {title} {A quantum
  mechanical open system as a model of a heat engine},}\ }\href
  {https://doi.org/10.1063/1.446862} {\bibfield  {journal} {\bibinfo  {journal}
  {The Journal of Chemical Physics}\ }\textbf {\bibinfo {volume} {80}},\
  \bibinfo {pages} {1625--1631} (\bibinfo {year} {1984})}\BibitemShut {NoStop}%
\bibitem [{\citenamefont {Geva}\ and\ \citenamefont
  {Kosloff}(1994)}]{geva_three-level_1994}%
  \BibitemOpen
  \bibfield  {author} {\bibinfo {author} {\bibfnamefont {E.}~\bibnamefont
  {Geva}}\ and\ \bibinfo {author} {\bibfnamefont {R.}~\bibnamefont {Kosloff}},\
  }\bibfield  {title} {\enquote {\bibinfo {title} {Three-level quantum
  amplifier as a heat engine: {{A}} study in finite-time thermodynamics},}\
  }\href {https://doi.org/10.1103/PhysRevE.49.3903} {\bibfield  {journal}
  {\bibinfo  {journal} {Phys. Rev. E}\ }\textbf {\bibinfo {volume} {49}},\
  \bibinfo {pages} {3903--3918} (\bibinfo {year} {1994})}\BibitemShut {NoStop}%
\bibitem [{\citenamefont {Geva}\ and\ \citenamefont
  {Kosloff}(1996)}]{geva_quantum_1996}%
  \BibitemOpen
  \bibfield  {author} {\bibinfo {author} {\bibfnamefont {E.}~\bibnamefont
  {Geva}}\ and\ \bibinfo {author} {\bibfnamefont {R.}~\bibnamefont {Kosloff}},\
  }\bibfield  {title} {\enquote {\bibinfo {title} {The quantum heat engine and
  heat pump: {{An}} irreversible thermodynamic analysis of the three-level
  amplifier},}\ }\href {https://doi.org/10.1063/1.471453} {\bibfield  {journal}
  {\bibinfo  {journal} {The Journal of Chemical Physics}\ }\textbf {\bibinfo
  {volume} {104}},\ \bibinfo {pages} {7681--7699} (\bibinfo {year}
  {1996})}\BibitemShut {NoStop}%
\bibitem [{\citenamefont {Mitchison}(2019)}]{mitchison_quantum_2019}%
  \BibitemOpen
  \bibfield  {author} {\bibinfo {author} {\bibfnamefont {M.~T.}\ \bibnamefont
  {Mitchison}},\ }\bibfield  {title} {\enquote {\bibinfo {title} {Quantum
  thermal absorption machines: Refrigerators, engines and clocks},}\ }\href
  {https://doi.org/10.1080/00107514.2019.1631555} {\bibfield  {journal}
  {\bibinfo  {journal} {Contemporary Physics}\ }\textbf {\bibinfo {volume}
  {60}},\ \bibinfo {pages} {164--187} (\bibinfo {year} {2019})}\BibitemShut
  {NoStop}%
\bibitem [{\citenamefont {Kavokin}\ and\ \citenamefont
  {Malpuech}(2003)}]{kavokin_cavity_2003}%
  \BibitemOpen
  \bibfield  {author} {\bibinfo {author} {\bibfnamefont {A.}~\bibnamefont
  {Kavokin}}\ and\ \bibinfo {author} {\bibfnamefont {G.}~\bibnamefont
  {Malpuech}},\ }\href@noop {} {\emph {\bibinfo {title} {Cavity Polaritons}}},\
  \bibinfo {edition} {1st}\ ed.,\ \bibinfo {series} {Thin Films and
  Nanostructures}\ No.~\bibinfo {number} {32}\ (\bibinfo  {publisher}
  {{Elsevier, Acad. Press}},\ \bibinfo {address} {{Amsterdam}},\ \bibinfo
  {year} {2003})\BibitemShut {NoStop}%
\bibitem [{\citenamefont {Tassone}\ \emph {et~al.}(1997)\citenamefont
  {Tassone}, \citenamefont {Piermarocchi}, \citenamefont {Savona},
  \citenamefont {Quattropani},\ and\ \citenamefont
  {Schwendimann}}]{tassone_bottleneck_1997}%
  \BibitemOpen
  \bibfield  {author} {\bibinfo {author} {\bibfnamefont {F.}~\bibnamefont
  {Tassone}}, \bibinfo {author} {\bibfnamefont {C.}~\bibnamefont
  {Piermarocchi}}, \bibinfo {author} {\bibfnamefont {V.}~\bibnamefont
  {Savona}}, \bibinfo {author} {\bibfnamefont {A.}~\bibnamefont
  {Quattropani}},\ and\ \bibinfo {author} {\bibfnamefont {P.}~\bibnamefont
  {Schwendimann}},\ }\bibfield  {title} {\enquote {\bibinfo {title} {Bottleneck
  effects in the relaxation and photoluminescence of microcavity polaritons},}\
  }\href {https://doi.org/10.1103/PhysRevB.56.7554} {\bibfield  {journal}
  {\bibinfo  {journal} {Phys. Rev. B}\ }\textbf {\bibinfo {volume} {56}},\
  \bibinfo {pages} {7554--7563} (\bibinfo {year} {1997})}\BibitemShut {NoStop}%
\bibitem [{\citenamefont {Tassone}\ and\ \citenamefont
  {Yamamoto}(1999)}]{tassone_exciton-exciton_1999}%
  \BibitemOpen
  \bibfield  {author} {\bibinfo {author} {\bibfnamefont {F.}~\bibnamefont
  {Tassone}}\ and\ \bibinfo {author} {\bibfnamefont {Y.}~\bibnamefont
  {Yamamoto}},\ }\bibfield  {title} {\enquote {\bibinfo {title}
  {Exciton-exciton scattering dynamics in a semiconductor microcavity and
  stimulated scattering into polaritons},}\ }\href
  {https://doi.org/10.1103/PhysRevB.59.10830} {\bibfield  {journal} {\bibinfo
  {journal} {Phys. Rev. B}\ }\textbf {\bibinfo {volume} {59}},\ \bibinfo
  {pages} {10830--10842} (\bibinfo {year} {1999})}\BibitemShut {NoStop}%
\bibitem [{\citenamefont {Doan}\ \emph {et~al.}(2005)\citenamefont {Doan},
  \citenamefont {Cao}, \citenamefont {Tran~Thoai},\ and\ \citenamefont
  {Haug}}]{doan_condensation_2005}%
  \BibitemOpen
  \bibfield  {author} {\bibinfo {author} {\bibfnamefont {T.~D.}\ \bibnamefont
  {Doan}}, \bibinfo {author} {\bibfnamefont {H.~T.}\ \bibnamefont {Cao}},
  \bibinfo {author} {\bibfnamefont {D.~B.}\ \bibnamefont {Tran~Thoai}},\ and\
  \bibinfo {author} {\bibfnamefont {H.}~\bibnamefont {Haug}},\ }\bibfield
  {title} {\enquote {\bibinfo {title} {Condensation kinetics of microcavity
  polaritons with scattering by phonons and polaritons},}\ }\href
  {https://doi.org/10.1103/PhysRevB.72.085301} {\bibfield  {journal} {\bibinfo
  {journal} {Phys. Rev. B}\ }\textbf {\bibinfo {volume} {72}},\ \bibinfo
  {pages} {085301} (\bibinfo {year} {2005})}\BibitemShut {NoStop}%
\bibitem [{\citenamefont {Hartwell}(2008)}]{hartwell_well_2008}%
  \BibitemOpen
  \bibfield  {author} {\bibinfo {author} {\bibfnamefont {V.~E.}\ \bibnamefont
  {Hartwell}},\ }\emph {\bibinfo {title} {Well - Microcavity Structures
  Experiments and Numerical Simulations}},\ \href@noop {} {Ph.D. thesis}
  (\bibinfo {year} {2008})\BibitemShut {NoStop}%
\bibitem [{\citenamefont {Porras}\ \emph {et~al.}(2002)\citenamefont {Porras},
  \citenamefont {Ciuti}, \citenamefont {Baumberg},\ and\ \citenamefont
  {Tejedor}}]{porras_polariton_2002}%
  \BibitemOpen
  \bibfield  {author} {\bibinfo {author} {\bibfnamefont {D.}~\bibnamefont
  {Porras}}, \bibinfo {author} {\bibfnamefont {C.}~\bibnamefont {Ciuti}},
  \bibinfo {author} {\bibfnamefont {J.~J.}\ \bibnamefont {Baumberg}},\ and\
  \bibinfo {author} {\bibfnamefont {C.}~\bibnamefont {Tejedor}},\ }\bibfield
  {title} {\enquote {\bibinfo {title} {Polariton dynamics and {{Bose-Einstein}}
  condensation in semiconductor microcavities},}\ }\href
  {https://doi.org/10.1103/PhysRevB.66.085304} {\bibfield  {journal} {\bibinfo
  {journal} {Phys. Rev. B}\ }\textbf {\bibinfo {volume} {66}},\ \bibinfo
  {pages} {085304} (\bibinfo {year} {2002})},\ \Eprint
  {https://arxiv.org/abs/cond-mat/0206276} {arxiv:cond-mat/0206276}
  \BibitemShut {NoStop}%
\bibitem [{\citenamefont {Hanai}\ \emph {et~al.}(2019)\citenamefont {Hanai},
  \citenamefont {Edelman}, \citenamefont {Ohashi},\ and\ \citenamefont
  {Littlewood}}]{hanai2019non}%
  \BibitemOpen
  \bibfield  {author} {\bibinfo {author} {\bibfnamefont {R.}~\bibnamefont
  {Hanai}}, \bibinfo {author} {\bibfnamefont {A.}~\bibnamefont {Edelman}},
  \bibinfo {author} {\bibfnamefont {Y.}~\bibnamefont {Ohashi}},\ and\ \bibinfo
  {author} {\bibfnamefont {P.~B.}\ \bibnamefont {Littlewood}},\ }\bibfield
  {title} {\enquote {\bibinfo {title} {Non-hermitian phase transition from a
  polariton bose-einstein condensate to a photon laser},}\ }\href@noop {}
  {\bibfield  {journal} {\bibinfo  {journal} {Physical review letters}\
  }\textbf {\bibinfo {volume} {122}},\ \bibinfo {pages} {185301} (\bibinfo
  {year} {2019})}\BibitemShut {NoStop}%
\bibitem [{\citenamefont {Shishkov}\ \emph {et~al.}(2022)\citenamefont
  {Shishkov}, \citenamefont {Andrianov}, \citenamefont {Zasedatelev},
  \citenamefont {Lagoudakis},\ and\ \citenamefont
  {Lozovik}}]{shishkov_exact_2021}%
  \BibitemOpen
  \bibfield  {author} {\bibinfo {author} {\bibfnamefont {V.~Y.}\ \bibnamefont
  {Shishkov}}, \bibinfo {author} {\bibfnamefont {E.~S.}\ \bibnamefont
  {Andrianov}}, \bibinfo {author} {\bibfnamefont {A.~V.}\ \bibnamefont
  {Zasedatelev}}, \bibinfo {author} {\bibfnamefont {P.~G.}\ \bibnamefont
  {Lagoudakis}},\ and\ \bibinfo {author} {\bibfnamefont {Y.~E.}\ \bibnamefont
  {Lozovik}},\ }\bibfield  {title} {\enquote {\bibinfo {title} {Exact
  analytical solution for the density matrix of a nonequilibrium polariton
  bose-einstein condensate},}\ }\href
  {https://doi.org/10.1103/PhysRevLett.128.065301} {\bibfield  {journal}
  {\bibinfo  {journal} {Phys. Rev. Lett.}\ }\textbf {\bibinfo {volume} {128}},\
  \bibinfo {pages} {065301} (\bibinfo {year} {2022})}\BibitemShut {NoStop}%
\bibitem [{\citenamefont {Imamoglu}\ \emph {et~al.}(1996)\citenamefont
  {Imamoglu}, \citenamefont {Ram}, \citenamefont {Pau},\ and\ \citenamefont
  {Yamamoto}}]{imamoglu_nonequilibrium_1996}%
  \BibitemOpen
  \bibfield  {author} {\bibinfo {author} {\bibfnamefont {A.}~\bibnamefont
  {Imamoglu}}, \bibinfo {author} {\bibfnamefont {R.~J.}\ \bibnamefont {Ram}},
  \bibinfo {author} {\bibfnamefont {S.}~\bibnamefont {Pau}},\ and\ \bibinfo
  {author} {\bibfnamefont {Y.}~\bibnamefont {Yamamoto}},\ }\bibfield  {title}
  {\enquote {\bibinfo {title} {Nonequilibrium condensates and lasers without
  inversion: {{Exciton-polariton}} lasers},}\ }\href
  {https://doi.org/10.1103/PhysRevA.53.4250} {\bibfield  {journal} {\bibinfo
  {journal} {Phys. Rev. A}\ }\textbf {\bibinfo {volume} {53}},\ \bibinfo
  {pages} {4250--4253} (\bibinfo {year} {1996})}\BibitemShut {NoStop}%
\bibitem [{\citenamefont {Kasprzak}\ \emph {et~al.}(2006)\citenamefont
  {Kasprzak}, \citenamefont {Richard}, \citenamefont {Kundermann},
  \citenamefont {Baas}, \citenamefont {Jeambrun}, \citenamefont {Keeling},
  \citenamefont {Marchetti}, \citenamefont {Szyma{\'n}ska}, \citenamefont
  {Andr{\'e}}, \citenamefont {Staehli}, \citenamefont {Savona}, \citenamefont
  {Littlewood}, \citenamefont {Deveaud},\ and\ \citenamefont
  {Dang}}]{kasprzak_boseeinstein_2006}%
  \BibitemOpen
  \bibfield  {author} {\bibinfo {author} {\bibfnamefont {J.}~\bibnamefont
  {Kasprzak}}, \bibinfo {author} {\bibfnamefont {M.}~\bibnamefont {Richard}},
  \bibinfo {author} {\bibfnamefont {S.}~\bibnamefont {Kundermann}}, \bibinfo
  {author} {\bibfnamefont {A.}~\bibnamefont {Baas}}, \bibinfo {author}
  {\bibfnamefont {P.}~\bibnamefont {Jeambrun}}, \bibinfo {author}
  {\bibfnamefont {J.~M.~J.}\ \bibnamefont {Keeling}}, \bibinfo {author}
  {\bibfnamefont {F.~M.}\ \bibnamefont {Marchetti}}, \bibinfo {author}
  {\bibfnamefont {M.~H.}\ \bibnamefont {Szyma{\'n}ska}}, \bibinfo {author}
  {\bibfnamefont {R.}~\bibnamefont {Andr{\'e}}}, \bibinfo {author}
  {\bibfnamefont {J.~L.}\ \bibnamefont {Staehli}}, \bibinfo {author}
  {\bibfnamefont {V.}~\bibnamefont {Savona}}, \bibinfo {author} {\bibfnamefont
  {P.~B.}\ \bibnamefont {Littlewood}}, \bibinfo {author} {\bibfnamefont
  {B.}~\bibnamefont {Deveaud}},\ and\ \bibinfo {author} {\bibfnamefont {L.~S.}\
  \bibnamefont {Dang}},\ }\bibfield  {title} {\enquote {\bibinfo {title}
  {Bose--{{Einstein}} condensation of exciton polaritons},}\ }\href
  {https://doi.org/10.1038/nature05131} {\bibfield  {journal} {\bibinfo
  {journal} {Nature}\ }\textbf {\bibinfo {volume} {443}},\ \bibinfo {pages}
  {409--414} (\bibinfo {year} {2006})}\BibitemShut {NoStop}%
\bibitem [{\citenamefont {Balili}\ \emph {et~al.}(2007)\citenamefont {Balili},
  \citenamefont {Hartwell}, \citenamefont {Snoke}, \citenamefont {Pfeiffer},\
  and\ \citenamefont {West}}]{balili_bose-einstein_2007}%
  \BibitemOpen
  \bibfield  {author} {\bibinfo {author} {\bibfnamefont {R.}~\bibnamefont
  {Balili}}, \bibinfo {author} {\bibfnamefont {V.}~\bibnamefont {Hartwell}},
  \bibinfo {author} {\bibfnamefont {D.}~\bibnamefont {Snoke}}, \bibinfo
  {author} {\bibfnamefont {L.}~\bibnamefont {Pfeiffer}},\ and\ \bibinfo
  {author} {\bibfnamefont {K.}~\bibnamefont {West}},\ }\bibfield  {title}
  {\enquote {\bibinfo {title} {Bose-{{Einstein Condensation}} of {{Microcavity
  Polaritons}} in a {{Trap}}},}\ }\href
  {https://doi.org/10.1126/science.1140990} {\bibfield  {journal} {\bibinfo
  {journal} {Science}\ }\textbf {\bibinfo {volume} {316}},\ \bibinfo {pages}
  {1007--1010} (\bibinfo {year} {2007})}\BibitemShut {NoStop}%
\bibitem [{\citenamefont {Kavokin}(2017)}]{kavokin_microcavities_2017}%
  \BibitemOpen
  \bibinfo {editor} {\bibfnamefont {A.}~\bibnamefont {Kavokin}},\ ed.,\
  \href@noop {} {\emph {\bibinfo {title} {Microcavities}}},\ \bibinfo {edition}
  {second edition}\ ed.,\ \bibinfo {series} {Series on Semiconductor Science
  and Technology}\ No.~\bibinfo {number} {16}\ (\bibinfo  {publisher} {{Oxford
  University Press}},\ \bibinfo {address} {{Oxford ; New York, NY}},\ \bibinfo
  {year} {2017})\BibitemShut {NoStop}%
\bibitem [{\citenamefont {Carusotto}\ and\ \citenamefont
  {Ciuti}(2013)}]{carusotto_quantum_2013}%
  \BibitemOpen
  \bibfield  {author} {\bibinfo {author} {\bibfnamefont {I.}~\bibnamefont
  {Carusotto}}\ and\ \bibinfo {author} {\bibfnamefont {C.}~\bibnamefont
  {Ciuti}},\ }\bibfield  {title} {\enquote {\bibinfo {title} {Quantum fluids of
  light},}\ }\href {https://doi.org/10.1103/RevModPhys.85.299} {\bibfield
  {journal} {\bibinfo  {journal} {Rev. Mod. Phys.}\ }\textbf {\bibinfo {volume}
  {85}},\ \bibinfo {pages} {299--366} (\bibinfo {year} {2013})}\BibitemShut
  {NoStop}%
\bibitem [{\citenamefont {Cao}\ \emph {et~al.}(2004)\citenamefont {Cao},
  \citenamefont {Doan}, \citenamefont {Tran~Thoai},\ and\ \citenamefont
  {Haug}}]{cao_condensation_2004}%
  \BibitemOpen
  \bibfield  {author} {\bibinfo {author} {\bibfnamefont {H.~T.}\ \bibnamefont
  {Cao}}, \bibinfo {author} {\bibfnamefont {T.~D.}\ \bibnamefont {Doan}},
  \bibinfo {author} {\bibfnamefont {D.~B.}\ \bibnamefont {Tran~Thoai}},\ and\
  \bibinfo {author} {\bibfnamefont {H.}~\bibnamefont {Haug}},\ }\bibfield
  {title} {\enquote {\bibinfo {title} {Condensation kinetics of cavity
  polaritons interacting with a thermal phonon bath},}\ }\href
  {https://doi.org/10.1103/PhysRevB.69.245325} {\bibfield  {journal} {\bibinfo
  {journal} {Phys. Rev. B}\ }\textbf {\bibinfo {volume} {69}},\ \bibinfo
  {pages} {245325} (\bibinfo {year} {2004})}\BibitemShut {NoStop}%
\bibitem [{\citenamefont {Kasprzak}\ \emph {et~al.}(2008)\citenamefont
  {Kasprzak}, \citenamefont {Solnyshkov}, \citenamefont {Andr{\'e}},
  \citenamefont {Dang},\ and\ \citenamefont
  {Malpuech}}]{kasprzak_formation_2008}%
  \BibitemOpen
  \bibfield  {author} {\bibinfo {author} {\bibfnamefont {J.}~\bibnamefont
  {Kasprzak}}, \bibinfo {author} {\bibfnamefont {D.~D.}\ \bibnamefont
  {Solnyshkov}}, \bibinfo {author} {\bibfnamefont {R.}~\bibnamefont
  {Andr{\'e}}}, \bibinfo {author} {\bibfnamefont {L.~S.}\ \bibnamefont
  {Dang}},\ and\ \bibinfo {author} {\bibfnamefont {G.}~\bibnamefont
  {Malpuech}},\ }\bibfield  {title} {\enquote {\bibinfo {title} {Formation of
  an {{Exciton Polariton Condensate}}: {{Thermodynamic}} versus {{Kinetic
  Regimes}}},}\ }\href {https://doi.org/10.1103/PhysRevLett.101.146404}
  {\bibfield  {journal} {\bibinfo  {journal} {Phys. Rev. Lett.}\ }\textbf
  {\bibinfo {volume} {101}},\ \bibinfo {pages} {146404} (\bibinfo {year}
  {2008})}\BibitemShut {NoStop}%
\bibitem [{\citenamefont {Piermarocchi}\ \emph {et~al.}(1996)\citenamefont
  {Piermarocchi}, \citenamefont {Tassone}, \citenamefont {Savona},
  \citenamefont {Quattropani},\ and\ \citenamefont
  {Schwendimann}}]{piermarocchi_nonequilibrium_1996}%
  \BibitemOpen
  \bibfield  {author} {\bibinfo {author} {\bibfnamefont {C.}~\bibnamefont
  {Piermarocchi}}, \bibinfo {author} {\bibfnamefont {F.}~\bibnamefont
  {Tassone}}, \bibinfo {author} {\bibfnamefont {V.}~\bibnamefont {Savona}},
  \bibinfo {author} {\bibfnamefont {A.}~\bibnamefont {Quattropani}},\ and\
  \bibinfo {author} {\bibfnamefont {P.}~\bibnamefont {Schwendimann}},\
  }\bibfield  {title} {\enquote {\bibinfo {title} {Nonequilibrium dynamics of
  free quantum-well excitons in time-resolved photoluminescence},}\ }\href
  {https://doi.org/10.1103/PhysRevB.53.15834} {\bibfield  {journal} {\bibinfo
  {journal} {Phys. Rev. B}\ }\textbf {\bibinfo {volume} {53}},\ \bibinfo
  {pages} {15834--15841} (\bibinfo {year} {1996})}\BibitemShut {NoStop}%
\bibitem [{\citenamefont {Hartwell}\ and\ \citenamefont
  {Snoke}(2010)}]{hartwell_numerical_2010}%
  \BibitemOpen
  \bibfield  {author} {\bibinfo {author} {\bibfnamefont {V.~E.}\ \bibnamefont
  {Hartwell}}\ and\ \bibinfo {author} {\bibfnamefont {D.~W.}\ \bibnamefont
  {Snoke}},\ }\bibfield  {title} {\enquote {\bibinfo {title} {Numerical
  simulations of the polariton kinetic energy distribution in {{GaAs}}
  quantum-well microcavity structures},}\ }\href
  {https://doi.org/10.1103/PhysRevB.82.075307} {\bibfield  {journal} {\bibinfo
  {journal} {Phys. Rev. B}\ }\textbf {\bibinfo {volume} {82}},\ \bibinfo
  {pages} {075307} (\bibinfo {year} {2010})}\BibitemShut {NoStop}%
\bibitem [{\citenamefont {Malpuech}\ \emph {et~al.}(2002)\citenamefont
  {Malpuech}, \citenamefont {Kavokin}, \citenamefont {Di~Carlo},\ and\
  \citenamefont {Baumberg}}]{malpuech_polariton_2002}%
  \BibitemOpen
  \bibfield  {author} {\bibinfo {author} {\bibfnamefont {G.}~\bibnamefont
  {Malpuech}}, \bibinfo {author} {\bibfnamefont {A.}~\bibnamefont {Kavokin}},
  \bibinfo {author} {\bibfnamefont {A.}~\bibnamefont {Di~Carlo}},\ and\
  \bibinfo {author} {\bibfnamefont {J.~J.}\ \bibnamefont {Baumberg}},\
  }\bibfield  {title} {\enquote {\bibinfo {title} {Polariton lasing by
  exciton-electron scattering in semiconductor microcavities},}\ }\href
  {https://doi.org/10.1103/PhysRevB.65.153310} {\bibfield  {journal} {\bibinfo
  {journal} {Phys. Rev. B}\ }\textbf {\bibinfo {volume} {65}},\ \bibinfo
  {pages} {153310} (\bibinfo {year} {2002})}\BibitemShut {NoStop}%
\bibitem [{\citenamefont {Baumberg}\ \emph {et~al.}(2002)\citenamefont
  {Baumberg}, \citenamefont {Savvidis}, \citenamefont {Lagoudakis},
  \citenamefont {Martin}, \citenamefont {Whittaker}, \citenamefont {Butte},
  \citenamefont {Skolnick},\ and\ \citenamefont
  {Roberts}}]{baumberg_polariton_2002}%
  \BibitemOpen
  \bibfield  {author} {\bibinfo {author} {\bibfnamefont {J.~J.}\ \bibnamefont
  {Baumberg}}, \bibinfo {author} {\bibfnamefont {P.~G.}\ \bibnamefont
  {Savvidis}}, \bibinfo {author} {\bibfnamefont {P.}~\bibnamefont
  {Lagoudakis}}, \bibinfo {author} {\bibfnamefont {M.}~\bibnamefont {Martin}},
  \bibinfo {author} {\bibfnamefont {D.}~\bibnamefont {Whittaker}}, \bibinfo
  {author} {\bibfnamefont {R.}~\bibnamefont {Butte}}, \bibinfo {author}
  {\bibfnamefont {M.}~\bibnamefont {Skolnick}},\ and\ \bibinfo {author}
  {\bibfnamefont {J.}~\bibnamefont {Roberts}},\ }\bibfield  {title} {\enquote
  {\bibinfo {title} {Polariton traps in semiconductor microcavities},}\ }\href
  {https://doi.org/10.1016/S1386-9477(02)00146-7} {\bibfield  {journal}
  {\bibinfo  {journal} {Physica E: Low-dimensional Systems and Nanostructures}\
  }\textbf {\bibinfo {volume} {13}},\ \bibinfo {pages} {385--389} (\bibinfo
  {year} {2002})}\BibitemShut {NoStop}%
\bibitem [{\citenamefont {Linden}, \citenamefont {Popescu},\ and\ \citenamefont
  {Skrzypczyk}(2010)}]{linden_how_2010}%
  \BibitemOpen
  \bibfield  {author} {\bibinfo {author} {\bibfnamefont {N.}~\bibnamefont
  {Linden}}, \bibinfo {author} {\bibfnamefont {S.}~\bibnamefont {Popescu}},\
  and\ \bibinfo {author} {\bibfnamefont {P.}~\bibnamefont {Skrzypczyk}},\
  }\bibfield  {title} {\enquote {\bibinfo {title} {How {{Small Can Thermal
  Machines Be}}? {{The Smallest Possible Refrigerator}}},}\ }\href
  {https://doi.org/10.1103/PhysRevLett.105.130401} {\bibfield  {journal}
  {\bibinfo  {journal} {Phys. Rev. Lett.}\ }\textbf {\bibinfo {volume} {105}},\
  \bibinfo {pages} {130401} (\bibinfo {year} {2010})}\BibitemShut {NoStop}%
\bibitem [{\citenamefont {Adesso}\ \emph {et~al.}(2018)\citenamefont {Adesso},
  \citenamefont {Anders}, \citenamefont {Binder}, \citenamefont {Correa},\ and\
  \citenamefont {Gogolin}}]{adesso_thermodynamics_2018}%
  \BibitemOpen
  \bibinfo {editor} {\bibfnamefont {G.}~\bibnamefont {Adesso}}, \bibinfo
  {editor} {\bibfnamefont {J.}~\bibnamefont {Anders}}, \bibinfo {editor}
  {\bibfnamefont {F.}~\bibnamefont {Binder}}, \bibinfo {editor} {\bibfnamefont
  {L.~A.}\ \bibnamefont {Correa}},\ and\ \bibinfo {editor} {\bibfnamefont
  {C.}~\bibnamefont {Gogolin}},\ eds.,\ \href
  {https://doi.org/10.1007/978-3-319-99046-0} {\emph {\bibinfo {title}
  {Thermodynamics in the {{Quantum Regime}}: {{Fundamental Aspects}} and {{New
  Directions}}}}},\ \bibinfo {edition} {1st}\ ed.,\ \bibinfo {series}
  {Fundamental {{Theories}} of {{Physics}}}\ No.\ \bibinfo {number} {195}\
  (\bibinfo  {publisher} {{Springer International Publishing : Imprint:
  Springer}},\ \bibinfo {address} {{Cham}},\ \bibinfo {year}
  {2018})\BibitemShut {NoStop}%
\bibitem [{\citenamefont {Murphy}, \citenamefont {Toledo~Tude},\ and\
  \citenamefont {Eastham}(2022)}]{murphy_laser_2022}%
  \BibitemOpen
  \bibfield  {author} {\bibinfo {author} {\bibfnamefont {C.~N.}\ \bibnamefont
  {Murphy}}, \bibinfo {author} {\bibfnamefont {L.}~\bibnamefont
  {Toledo~Tude}},\ and\ \bibinfo {author} {\bibfnamefont {P.~R.}\ \bibnamefont
  {Eastham}},\ }\bibfield  {title} {\enquote {\bibinfo {title} {Laser {{Cooling
  Beyond Rate Equations}}: {{Approaches}} from {{Quantum Thermodynamics}}},}\
  }\href {https://doi.org/10.3390/app12031620} {\bibfield  {journal} {\bibinfo
  {journal} {Appl. Sci.}\ }\textbf {\bibinfo {volume} {12}},\ \bibinfo {pages}
  {1620} (\bibinfo {year} {2022})}\BibitemShut {NoStop}%
\bibitem [{\citenamefont {Esposito}, \citenamefont {Harbola},\ and\
  \citenamefont {Mukamel}(2009)}]{esposito_nonequilibrium_2009}%
  \BibitemOpen
  \bibfield  {author} {\bibinfo {author} {\bibfnamefont {M.}~\bibnamefont
  {Esposito}}, \bibinfo {author} {\bibfnamefont {U.}~\bibnamefont {Harbola}},\
  and\ \bibinfo {author} {\bibfnamefont {S.}~\bibnamefont {Mukamel}},\
  }\bibfield  {title} {\enquote {\bibinfo {title} {Nonequilibrium fluctuations,
  fluctuation theorems, and counting statistics in quantum systems},}\ }\href
  {https://doi.org/10.1103/RevModPhys.81.1665} {\bibfield  {journal} {\bibinfo
  {journal} {Rev. Mod. Phys.}\ }\textbf {\bibinfo {volume} {81}},\ \bibinfo
  {pages} {1665--1702} (\bibinfo {year} {2009})}\BibitemShut {NoStop}%
\bibitem [{\citenamefont {Gasparinetti}\ \emph {et~al.}(2014)\citenamefont
  {Gasparinetti}, \citenamefont {Solinas}, \citenamefont {Braggio},\ and\
  \citenamefont {Sassetti}}]{gasparinetti_heat-exchange_2014}%
  \BibitemOpen
  \bibfield  {author} {\bibinfo {author} {\bibfnamefont {S.}~\bibnamefont
  {Gasparinetti}}, \bibinfo {author} {\bibfnamefont {P.}~\bibnamefont
  {Solinas}}, \bibinfo {author} {\bibfnamefont {A.}~\bibnamefont {Braggio}},\
  and\ \bibinfo {author} {\bibfnamefont {M.}~\bibnamefont {Sassetti}},\
  }\bibfield  {title} {\enquote {\bibinfo {title} {Heat-exchange statistics in
  driven open quantum systems},}\ }\href
  {https://doi.org/10.1088/1367-2630/16/11/115001} {\bibfield  {journal}
  {\bibinfo  {journal} {New J. Phys.}\ }\textbf {\bibinfo {volume} {16}},\
  \bibinfo {pages} {115001} (\bibinfo {year} {2014})}\BibitemShut {NoStop}%
\bibitem [{\citenamefont {Li}\ \emph {et~al.}(2017)\citenamefont {Li},
  \citenamefont {Kim}, \citenamefont {Agarwal},\ and\ \citenamefont
  {Scully}}]{li_quantum_2017}%
  \BibitemOpen
  \bibfield  {author} {\bibinfo {author} {\bibfnamefont {S.-W.}\ \bibnamefont
  {Li}}, \bibinfo {author} {\bibfnamefont {M.~B.}\ \bibnamefont {Kim}},
  \bibinfo {author} {\bibfnamefont {G.~S.}\ \bibnamefont {Agarwal}},\ and\
  \bibinfo {author} {\bibfnamefont {M.~O.}\ \bibnamefont {Scully}},\ }\bibfield
   {title} {\enquote {\bibinfo {title} {Quantum statistics of a single-atom
  {{Scovil-Schulz-DuBois}} heat engine},}\ }\href
  {https://doi.org/10.1103/PhysRevA.96.063806} {\bibfield  {journal} {\bibinfo
  {journal} {Phys. Rev. A}\ }\textbf {\bibinfo {volume} {96}},\ \bibinfo
  {pages} {063806} (\bibinfo {year} {2017})}\BibitemShut {NoStop}%
\bibitem [{\citenamefont {Busley}\ \emph {et~al.}(2023)\citenamefont {Busley},
  \citenamefont {Espert~Miranda}, \citenamefont {Kurtscheid}, \citenamefont
  {Wolf}, \citenamefont {Vewinger}, \citenamefont {Schmitt},\ and\
  \citenamefont {Weitz}}]{busley_sunlight-pumped_2023}%
  \BibitemOpen
  \bibfield  {author} {\bibinfo {author} {\bibfnamefont {E.}~\bibnamefont
  {Busley}}, \bibinfo {author} {\bibfnamefont {L.}~\bibnamefont
  {Espert~Miranda}}, \bibinfo {author} {\bibfnamefont {C.}~\bibnamefont
  {Kurtscheid}}, \bibinfo {author} {\bibfnamefont {F.}~\bibnamefont {Wolf}},
  \bibinfo {author} {\bibfnamefont {F.}~\bibnamefont {Vewinger}}, \bibinfo
  {author} {\bibfnamefont {J.}~\bibnamefont {Schmitt}},\ and\ \bibinfo {author}
  {\bibfnamefont {M.}~\bibnamefont {Weitz}},\ }\bibfield  {title} {\enquote
  {\bibinfo {title} {Sunlight-pumped two-dimensional thermalized photon gas},}\
  }\href {https://doi.org/10.1103/PhysRevA.107.052204} {\bibfield  {journal}
  {\bibinfo  {journal} {Phys. Rev. A}\ }\textbf {\bibinfo {volume} {107}},\
  \bibinfo {pages} {052204} (\bibinfo {year} {2023})}\BibitemShut {NoStop}%
\bibitem [{\citenamefont {Klembt}\ \emph {et~al.}(2015)\citenamefont {Klembt},
  \citenamefont {Durupt}, \citenamefont {Datta}, \citenamefont {Klein},
  \citenamefont {Baas}, \citenamefont {L{\'e}ger}, \citenamefont {Kruse},
  \citenamefont {Hommel}, \citenamefont {Minguzzi},\ and\ \citenamefont
  {Richard}}]{klembt_exciton-polariton_2015}%
  \BibitemOpen
  \bibfield  {author} {\bibinfo {author} {\bibfnamefont {S.}~\bibnamefont
  {Klembt}}, \bibinfo {author} {\bibfnamefont {E.}~\bibnamefont {Durupt}},
  \bibinfo {author} {\bibfnamefont {S.}~\bibnamefont {Datta}}, \bibinfo
  {author} {\bibfnamefont {T.}~\bibnamefont {Klein}}, \bibinfo {author}
  {\bibfnamefont {A.}~\bibnamefont {Baas}}, \bibinfo {author} {\bibfnamefont
  {Y.}~\bibnamefont {L{\'e}ger}}, \bibinfo {author} {\bibfnamefont
  {C.}~\bibnamefont {Kruse}}, \bibinfo {author} {\bibfnamefont
  {D.}~\bibnamefont {Hommel}}, \bibinfo {author} {\bibfnamefont
  {A.}~\bibnamefont {Minguzzi}},\ and\ \bibinfo {author} {\bibfnamefont
  {M.}~\bibnamefont {Richard}},\ }\bibfield  {title} {\enquote {\bibinfo
  {title} {Exciton-{{Polariton Gas}} as a {{Nonequilibrium Coolant}}},}\ }\href
  {https://doi.org/10.1103/PhysRevLett.114.186403} {\bibfield  {journal}
  {\bibinfo  {journal} {Phys. Rev. Lett.}\ }\textbf {\bibinfo {volume} {114}},\
  \bibinfo {pages} {186403} (\bibinfo {year} {2015})}\BibitemShut {NoStop}%
\end{thebibliography}%
\end{document}